\documentclass[12pt]{article}

\usepackage{scicite}

\usepackage{times}
\usepackage{graphicx}

\newenvironment{sciabstract}{%
\begin{quote} \bf}
{\end{quote}}

\newcounter{lastnote}

\title{Detection of an oxygen emission line from a high redshift galaxy
in the reionization epoch}

\author
{Akio K. Inoue$^{1\ast}$, Yoichi Tamura$^{2}$, Hiroshi Matsuo$^{3,4}$, 
Ken Mawatari$^{1}$, Ikkoh Shimizu$^{5}$, \\ Takatoshi Shibuya$^{6}$,
Kazuaki Ota$^{7,8}$, Naoki Yoshida$^{9,10}$, Erik Zackrisson$^{11}$, \\
Nobunari Kashikawa$^{3,4}$, Kotaro Kohno$^{2}$, Hideki Umehata$^{2,12}$,
Bunyo Hatsukade$^{3}$, \\ Masanori Iye$^{3}$, Yuichi Matsuda$^{3,4}$, 
Takashi Okamoto$^{13}$, Yuki Yamaguchi$^{2}$\\
\\
\normalsize{$^{1}$College of General Education, Osaka Sangyo
University,}\\
\normalsize{3-1-1 Nakagaito, Daito, Osaka 574-8530, Japan}\\ 
\normalsize{$^{2}$Institute of Astronomy, University of Tokyo, 2-21-1
Osawa, Mitaka, Tokyo 181-0015, Japan}\\
\normalsize{$^{3}$National Astronomical Observatory of Japan, 2-21-1
Osawa, Mitaka, Tokyo 181-8588, Japan}\\
\normalsize{$^{4}$The Graduate University for Advanced Studies
(SOKENDAI),}\\
\normalsize{2-21-1 Osawa, Mitaka, Tokyo 181-8588, Japan}\\
\normalsize{$^{5}$Department of Earth and Space Science, Osaka
University,}\\
\normalsize{1-1 Machikaneyama, Toyonaka, Osaka 560-0043, Japan}\\
\normalsize{$^{6}$Institute for Cosmic Ray Research, University of
Tokyo,}\\
\normalsize{5-1-5 Kashiwanoha, Kashiwa, Chiba 277-8582, Japan}\\
\normalsize{$^{7}$Kavli Institute for Cosmology, University of Cambridge,
Madingley Road, Cambridge CB3 0HA, UK}\\
\normalsize{$^{8}$Cavendish Laboratory, University of Cambridge, 19
J.J. Thomson Avenue, Cambridge CB3 0HE, UK}\\
\normalsize{$^{9}$Department of Physics, University of Tokyo, 7-3-1
Hongo, Bunkyo, Tokyo 113-0033, Japan}\\
\normalsize{$^{10}$Kavli IPMU, University of Tokyo, 5-1-5 Kashiwanoha,
Kashiwa, Chiba 277-8583, Japan}\\
\normalsize{$^{11}$Department of Physics and Astronomy, Uppsala
University, Box 515, SE-751 20 Uppsala, Sweden}\\
\normalsize{$^{12}$European Southern Observatory,
Karl-Schwarzschild-Str. 2, D-85748 Garching, Germany}\\
\normalsize{$^{13}$Department of Cosmosciences, Graduates School of
Science, Hokkaido University,}\\
\normalsize{N10 W8, Kitaku, Sapporo, Hokkaido 060-0810, Japan}\\
\\
\normalsize{$^\ast$To whom correspondence should be addressed; 
E-mail: akinoue@las.osaka-sandai.ac.jp}
}

\date{}

\begin{document} 


\baselineskip24pt


\maketitle 


\clearpage

\begin{sciabstract}
 The physical properties and elemental abundances of the interstellar
 medium in galaxies during cosmic reionization are important for 
 understanding the role of galaxies in this process. We report  
 the Atacama Large Millimeter/submillimeter Array detection of an
 oxygen emission line at a wavelength of 88 micrometers from a galaxy at an
 epoch about 700 million years after the Big Bang. The oxygen abundance
 of this galaxy is estimated at about one-tenth that of the Sun. The 
 non-detection of far-infrared continuum emission indicates a deficiency
 of interstellar dust in the galaxy. A carbon emission line at a
 wavelength of 158 micrometers is also not detected, implying an unusually
 small amount of neutral gas. These properties might allow ionizing
 photons to escape into the intergalactic medium.
\end{sciabstract}


The physical and chemical conditions of the interstellar medium
(ISM) in galaxies can be revealed with forbidden atomic emission lines
from the warm-phase ISM, such as ionized hydrogen (H~{\sc ii}) regions and
photodissociation regions (PDRs). A far-infrared (FIR) forbidden
emission line, the [C~{\sc ii}] 158 $\mu$m line predominantly coming
from PDRs, has already been detected in many high-$z$ objects
\cite{stacey10,gullberg15}. Recent observations with the Atacama
Large Millimeter/submillimeter Array (ALMA) have revealed the [C~{\sc
ii}] line emission from young star-forming galaxies emitting a strong
hydrogen Ly$\alpha$ line, so-called Ly$\alpha$ emitters (LAEs), at
redshift $z\sim5$ to 6 \cite{capak15,knudsen16}. However, ALMA 
observations have also shown that LAEs at $z>6$ have at least an order
of magnitude lower luminosity of the [C~{\sc ii}] line than that
expected from their star formation rate (SFR)
\cite{ouchi13,ota14,maiolino15,knudsen16},  
suggesting unusual ISM conditions in these high-$z$ LAEs
\cite{vallini15}.

{\it Herschel} observations of nearby dwarf galaxies, on the other hand,
have revealed that a forbidden oxygen line, [O~{\sc iii}] 88 $\mu$m, is
much stronger than the [C~{\sc ii}] line in these chemically unevolved
galaxies \cite{madden13,delooze14,cormier15}. The {\it Infrared
Space Observatory} and the Japanese infrared astronomical satellite,
AKARI, have detected the [O~{\sc iii}]  
line from the Large Magellanic Cloud and from many nearby galaxies
\cite{kawada11,brauher08}. However, the [O~{\sc iii}] line has
rarely been discussed in a high-$z$ context because of the lack of 
instruments suitable to observe the redshifted line. Only a few
detections from gravitationally lensed dusty starburst galaxies with
active galactic nuclei 
at $z\sim3$ to 4 have been reported \cite{ferkinhoff10,valtchanov11}
prior to ALMA. On the other hand, simulations predict that ALMA will
be able to detect the [O~{\sc iii}] line from star-forming galaxies
with reasonable integration time even at $z>8$ \cite{inoue14}.

To examine the [O~{\sc iii}] 88 $\mu$m line in high-$z$ LAEs,
we performed ALMA observations of an LAE at $z=7.2$, SXDF-NB1006-2,
discovered with the Subaru Telescope \cite{shibuya12}. We have also
obtained ALMA data of the [C~{\sc ii}] 158 $\mu$m line of this
galaxy. The observations and the data reduction are described in
\cite{som}. The [O~{\sc iii}] line is detected with a significance
of $5.3\sigma$ (Figure~1A) and the obtained line flux is
$6.2\times10^{-21}$ W m$^{-2}$; the corresponding luminosity is
$3.8\times10^{35}$ W (Table~1). The [C~{\sc ii}] line is not detected 
at the position of the [O~{\sc iii}] emission line and we take
the $3\sigma$ upper limit for the [C~{\sc ii}] line flux as
$<5.3\times10^{-22}$ W m$^{-2}$. However, we note a marginal signal
($3.5\sigma$) that displays a spatial offset ($\approx0.4''\approx2$ kpc
in the proper distance) from the [O~{\sc iii}] emission (fig.~S4).
The continuum is not detected in either of the ALMA bands, resulting in a
$3\sigma$ upper limit of the total IR luminosity of $<2.9\times10^{37}$
W when assuming a dust temperature of 40 K and an emissivity
index of 1.5.

The spatial distribution of the ALMA [O~{\sc iii}] emission overlaps
with that of the Subaru Ly$\alpha$ emission (Figure~1A) as expected
because both emission lines are produced in the same ionized gas. On the
other hand, the Ly$\alpha$ emission is well resolved (the image
resolution is $0.4''$) and spatially more extended than the 
[O~{\sc iii}] line. This is because Ly$\alpha$ photons suffer from
resonant scattering by neutral hydrogen atoms in the gas surrounding
the galaxy. The systemic redshift of the galaxy is estimated at
$z=7.2120\pm0.0003$ from the [O~{\sc iii}] emission line at
an observed wavelength of 725.603 $\mu$m. The Ly$\alpha$ line is
located at $\Delta v_{\rm Ly\alpha}=+1.1(\pm0.3)\times10^2$ km
s$^{-1}$ relative to the systemic redshift (Fig.~1C and fig.~S6). 
This velocity offset, caused by scattering of neutral hydrogen, is
relatively small by comparison to those observed in galaxies at $z\sim2$
to 3 ($\Delta v_{\rm Ly\alpha}\sim300$ km~s$^{-1}$), given the ultraviolet
(UV) absolute magnitude of this galaxy ($M_{\rm UV}=-21.53$ AB) 
\cite{hashimoto13,erb14,shibuya14b}. The observed small 
$\Delta v_{\rm Ly\alpha}$ of SXDF-NB1006-2 may indicate an H~{\sc i}
column density of $N_{\rm HI}<10^{20}$ cm$^{-2}$
\cite{shibuya14b,hashimoto15}. SXDF-NB1006-2 is in the reionization era 
where only the intergalactic medium (IGM) with a high hydrogen
neutral fraction may cause an observation of 
$\Delta v_{\rm Ly\alpha}\approx+100$ km s$^{-1}$ \cite{kakiichi15}, 
implying an even smaller 
H~{\sc i} column density in the ISM of this galaxy.

We performed spectral energy distribution (SED) modeling 
to derive physical quantities such as the SFR of SXDF-NB1006-2
(Table~1). In addition to broadband photometric data from the United
Kingdom Infra-Red Telescope (UKIRT) $J$, $H$, and $K$ bands, {\it
Spitzer} 3.6-$\mu$m and 4.5-$\mu$m bands and Subaru 
narrowband photometry $NB1006$ (table~S3), we have also used
the [O~{\sc iii}] line flux and the IR luminosity as constraints
(Fig.~2) \cite{som}. The extremely blue rest-frame UV color of this
galaxy [slope $\beta<-2.6$ ($3\sigma$) estimated from $J-H$, where the
flux density $F_\lambda\propto\lambda^\beta$], gives an age of
$\sim1$ million years for the ongoing star formation episode. The
non-detection of the dust IR emission suggests little dust and hence
negligible attenuation. The observed strong [O~{\sc iii}] line flux
favors an oxygen abundance of 5\% to 100\% that of the Sun but rejects 2\%
and 200\% of the solar abundance at $>95\%$ confidence. The
obtained oxygen abundance is similar to those estimated in galaxies at
$z\sim6$ to 7 for which UV C~{\sc iii}] and C~{\sc iv} emission lines were
detected \cite{stark15a,stark15b}. Because the $\sim1$ million years is
insufficient to produce the inferred oxygen abundance, the galaxy must have had
previous star formation episodes. Therefore, the derived stellar mass of
$\sim300$ million solar masses is regarded as a lower limit. We obtain a
$\sim50\%$ escape fraction of hydrogen-ionizing photons to the IGM in
the best-fit model. Such a high escape fraction, although still
uncertain, may imply a low H~{\sc i} column density of $\sim10^{17}$
cm$^{-2}$ \cite{fescnhi} or porous structure in the ISM of the galaxy.

The [O~{\sc iii}]/far-UV luminosity ratio of SXDF-NB1006-2 is
similar to those of nearby dwarf galaxies with an oxygen abundance of
10\% to 60\% that of the Sun (Fig.~3A), which suggests that the oxygen
abundance estimated from the SED modeling is reasonable and chemical
enrichment in this young galaxy has already proceeded. On the other
hand, the dust IR continuum and the [C~{\sc ii}] line of
SXDF-NB1006-2 are very weak relative to those of the nearby dwarf
galaxies (Fig.~3, B and C). The galaxies at $z\sim3$ to 4, from which
the [O~{\sc iii}] line was detected previously, are IR luminous dusty
ones \cite{ferkinhoff10,valtchanov11}. Their [O~{\sc iii}]/IR and
[O~{\sc iii}]/[C~{\sc ii}] luminosity ratios are similar to those of
nearby spiral galaxies \cite{delooze14} and are at least one order of
magnitude smaller than those of SXDF-NB1006-2. The high [O~{\sc
iii}]/IR ratio of SXDF-NB1006-2 despite a degree of chemical enrichment
(or so-called metallicity) similar to that of the nearby dwarf galaxies
indicates a very small mass fraction of dust in elements 
heavier than helium (or dust-to-metal mass ratio) in SXDF-NB1006-2. 
The dust deficiency of this galaxy is in contrast to the discovery of a
dusty galaxy at $z\approx7.5$ \cite{watson15}, suggesting a diversity
of the dust content in the reionization epoch. Because the 
[C~{\sc ii}] line predominantly arises in gas where hydrogen is neutral,
the non-detection of the [C~{\sc ii}] line in SXDF-NB1006-2 suggests
that this young galaxy has little H~{\sc i} gas.

We also compared the observed properties of SXDF-NB1006-2 with the
galaxies at $z=7.2$ in a cosmological hydrodynamic simulation of
galaxy formation and evolution \cite{som}. The simulation yielded
several galaxies with a UV luminosity similar to that of 
SXDF-NB1006-2 (fig.~S10). Relative to these simulated galaxies, 
SXDF-NB1006-2 has the highest [O~{\sc iii}] line luminosity, 
a similar oxygen abundance and a lower dust IR luminosity by at least a
factor of 2 to 3. This indicates a factor of $>2$ to 3 smaller 
dust-to-metal mass ratio within the ISM of SXDF-NB1006-2 relative to that
in the simulated galaxies where we assumed the dust-to-metal mass ratio
of 50\% as in the Milky Way ISM \cite{inoue11b}. Therefore, the
dust-to-metal ratio of SXDF-NB1006-2 is implied to be $<$20\%. 
The dust-to-metal ratio is determined by two processes: dust growth by
accretion of atoms and molecules onto the existing grains in cold dense
clouds, and dust destruction by supernova (SN) shock waves consequent
upon star formation \cite{inoue11b}. The dust-poor nature of
SXDF-NB1006-2 may be explained by rapid dust destruction due to its high
SN rate or by slow accretion growth due to a lack of cold dense clouds
in the ISM.

In the context of cosmic reionization studies, the most uncertain
parameter is the product of the escape fraction of ionizing photons and
the emission efficiency of these photons: $f_{\rm esc}\xi_{\rm ion}$
\cite{bouwens15b}. From the SED modeling, we have obtained 
$\log_{10}(f_{\rm esc}\xi_{\rm ion}[{\rm Hz~erg^{-1}}])=25.44^{+0.46}_{-0.84}$ 
for SXDF-NB1006-2 \cite{som}. This ionizing photon emission
efficiency is strong enough to reach (or even exceed) the cosmic
ionizing photon emissivity at $z\sim7$ estimated from various
observational constraints on reionization \cite{bouwens15b} by an
accumulation of galaxies that have already been detected ($M_{\rm
UV}<-17$), although this does not rule out contributions of fainter,
currently undetected galaxies to the ionizing emissivity. The ISM properties of
SXDF-NB1006-2, with little dust and H~{\sc i} gas, may make
this galaxy a prototypical example of a source of cosmic reionization.


\bibliography{z7O3C2}

\begin{thebibliography}{10}
\bibitem{fescnhi}
The photoionization cross section of a neutral hydrogen atom at the wavelength
  of 912 \AA\ (Lyman limit) is $\sigma=6.3\times10^{-18}$ cm$^2$. Then, the
  H~{\sc i} column density for a 50\% escape is $N_{\rm HI}=-\ln
  0.5/\sigma\simeq1\times10^{17}$ cm$^{-2}$ in a uniform medium.
\end{thebibliography}

\bibliographystyle{Science}



\paragraph*{Acknowledgements}

This paper makes use of the following ALMA data: ADS/JAO.ALMA\# 
2013.1.01010.S and 2012.1.00374.S which are available at 
https://almascience.nao.ac.jp/alma-data/archive.
ALMA is a partnership of the European Southern Observatory (ESO)
(representing its member states), NSF (USA) and NINS (Japan), together
with NRC (Canada), NSC and ASIAA (Taiwan), and KASI (Republic of Korea),
in cooperation with the Republic of Chile. The Joint ALMA Observatory is
operated by ESO, the National Radio Astronomy Observatory/Associated
Universities Inc., and the National Astronomical Observatory of Japan
(NAOJ). Based in part on data collected at Subaru Telescope, which is
operated by NAOJ; data are available at http://smoka.nao.ac.jp/ under
project codes S08B-019, S08B-051, and S09B-055, and also at the
W.M. Keck Observatory, which is operated as a scientific partnership
among the California Institute of Technology, the University of
California and NASA; data are available at 
www2.keck.hawaii.edu/koa/public/koa.php 
under the project code S331D. When some of the data reported here were
acquired, UKIRT was operated by the Joint Astronomy Centre on behalf of
the Science and Technology Facilities Council of the U.K.; UKIDSS data
are available at http://wsa.roe.ac.uk//dr10plus\_release.html.
Based in part on archival data obtained with the Spitzer Space
Telescope, which is operated by the Jet Propulsion Laboratory,
California Institute of Technology under a contract with NASA; data are 
available at www.cfa.harvard.edu/SEDS/data.html.

Supported by JSPS KAKENHI grant numbers 26287034
(A.K.I. and K.M.), 26247022 (I.S.), 25287050 (N.Y.), 24740112 (T.O.),
15H02073 (Y.T.) and 15K17616 (B.H.), and by a grant-in-aid for JSPS
Fellows (H.U.), by a grant-in-aid for the Global COE Program ``The
Next Generation of Physics, Spun from Universality and Emergence'' from
MEXT of Japan (K.O.), the Kavli Institute Fellowship at (Kavli
Institute for Cosmology, University of Cambridge) supported by the
Kavli Foundation (K.O.), and the Swedish Research Council (project
2011-5349) and the Wenner-Gren Foundations (E.Z.).

\vspace{1cm}

{\bf Supplementary Materials:}

Materials and Methods

Figures S1-S12

Tables S1-S4

References (31-75)


\clearpage

\begin{figure}[htbp]
\begin{center}
\includegraphics[width=0.45\textwidth,keepaspectratio,clip,angle=-90]{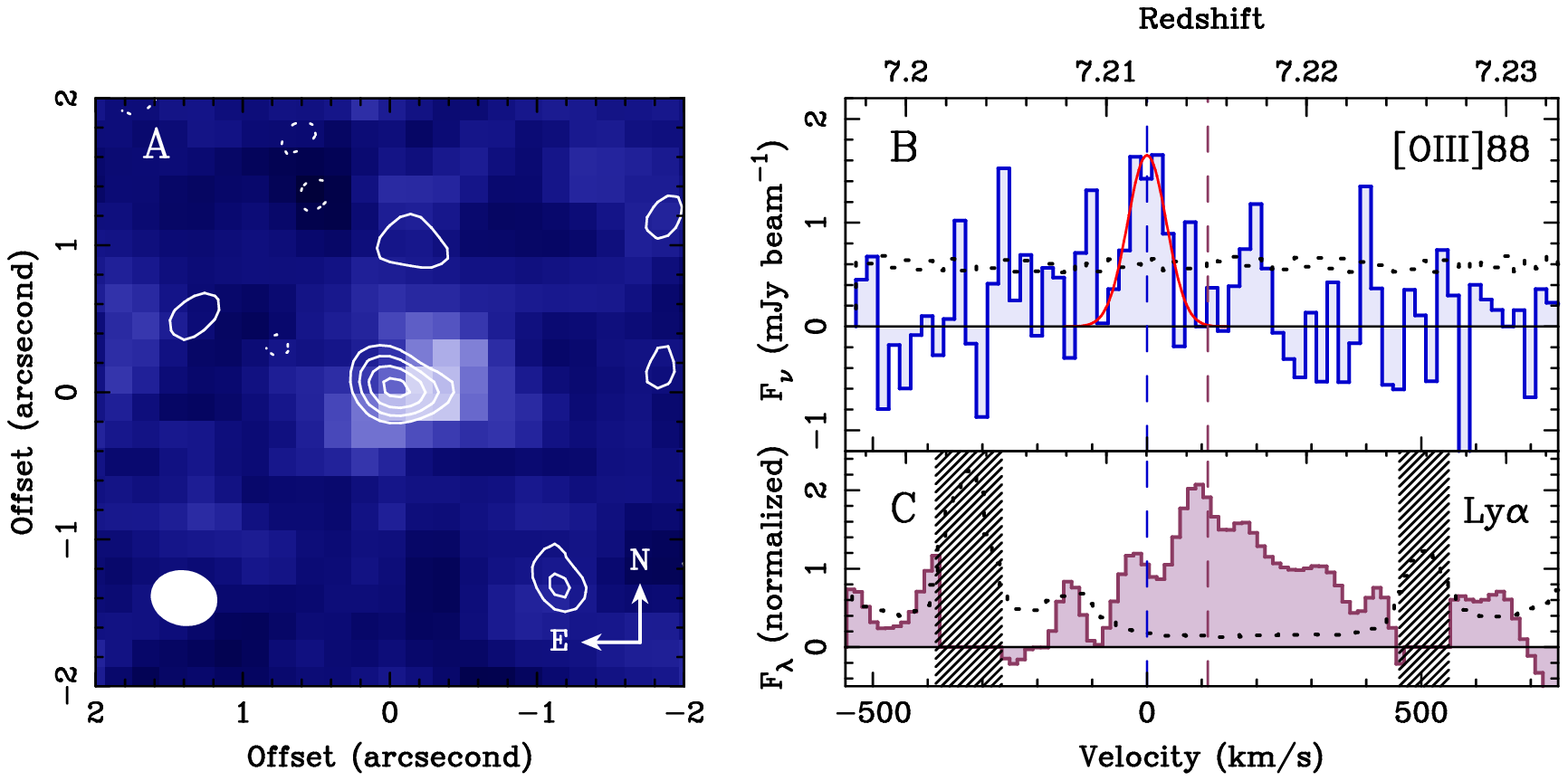}
\caption{\textsf{[O~\textsc{iii}] 88 $\mu$m and Ly$\alpha$ emission
 images and spectra of SXDF-NB1006-2.}
 (\textsf{A}) The ALMA [O~\textsc{iii}] 88 $\mu$m image (contours)
 overlaid on the Subaru narrow-band Ly$\alpha$ image (offsets from
 the position listed in Table 1). Contours are drawn at ($-$2, 2, 3, 4,
 5)$\times\sigma$, where $\sigma=0.0636$ Jy~beam$^{-1}$~km~s$^{-1}$. 
 Negative contours are shown by the dotted line. The ellipse at lower
 left represents the synthesized beam size of ALMA.
 (\textsf{B}) The ALMA [O~\textsc{iii}] 88 $\mu$m spectrum with
 resolution of 20 km~s$^{-1}$ at the intensity peak position shown against
 the relative velocity with respect to the redshift $z = 7.2120$ (blue
 dashed line). The best-fit Gaussian profile for the [O~{\sc iii}] line
 is overlaid. The RMS noise level is shown by the dotted line.
 (\textsf{C}) The Ly$\alpha$ spectrum \cite{shibuya12} shown as a
 function of the relative velocity compared to the [O~\textsc{iii}] 88
 $\mu$m line. The flux density is normalized by a unit of $10^{-18}$
 erg~s$^{-1}$~cm$^{-2}$~\AA$^{-1}$. The sky level on an arbitrary scale
 is shown by the dotted line. The velocity intervals where Earth's
 atmospheric lines severely contaminate the spectrum are flagged 
 (hatched boxes). The Ly$\alpha$ line shows a velocity shift 
 $\Delta v\approx+110$ km s$^{-1}$ relative to the [O~{\sc iii}] line
 (red dashed line).
}
\label{fig-OIII.tap0.3arcsec_x65y65_maintext}
\end{center}
\end{figure}

\clearpage

\begin{figure}
 \begin{center}
  \includegraphics[width=12cm]{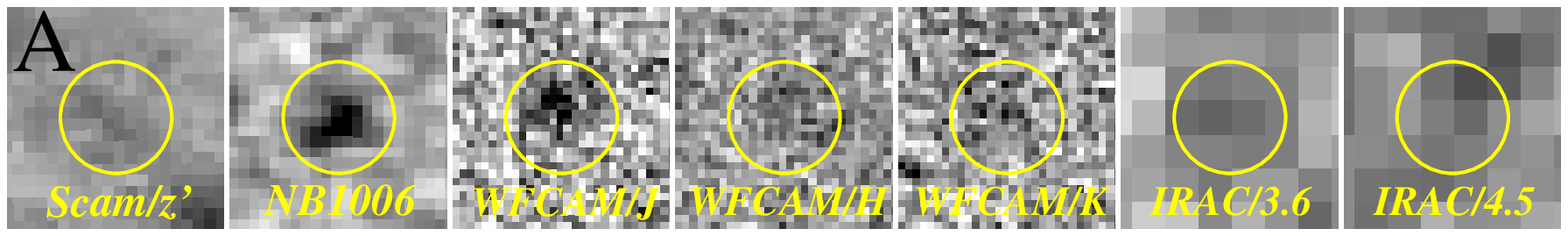}\\
  \includegraphics[width=12cm]{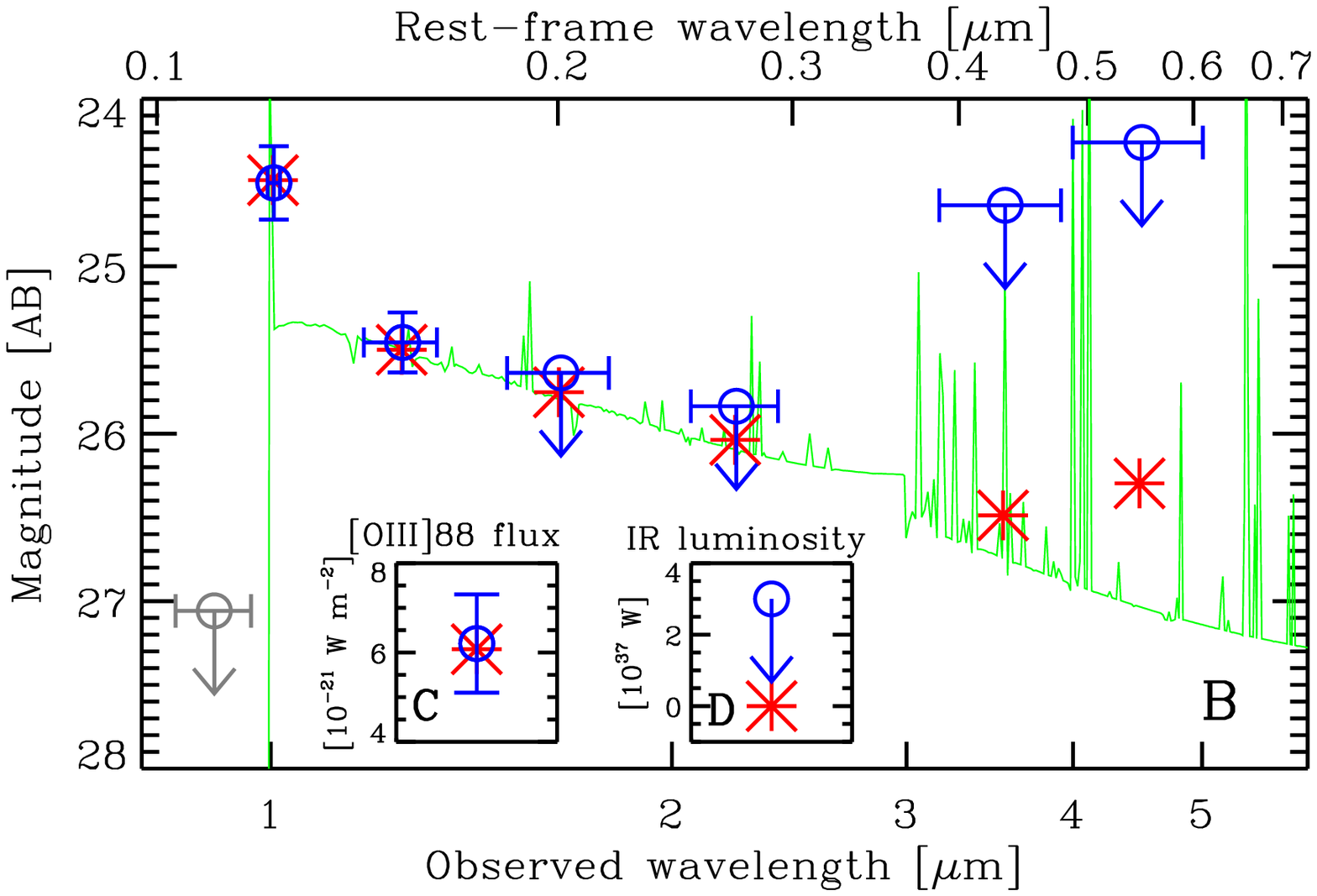}
  \caption{\textsf{Spectral energy distribution of SXDF-NB1006-2.} 
  (\textsf{A}) Thumbnail images ($4''\times4''$, north is up, east is to
  the left) in the Subaru/Suprime-Cam $z'$, $NB1006$, UKIRT/WFCAM $J$,
  $H$, and $K$ bands, and {\it Spitzer}/IRAC 3.6-$\mu$m and 4.5-$\mu$m
  bands, from left to right.
  (\textsf{B}) Near-infrared photometric data with the best-fit
  model. The bottom horizontal axis shows the wavelength in the
  observer's rest frame; the upper axis shows that in the source
  rest frame. The observations are marked by the circles. The
  horizontal error bars show the wavelength range of the band filters. 
  The vertical error bars for detection bands represent $\pm1\sigma$
  photometric uncertainties; the downward arrows for non-detections
  represent the $3\sigma$ upper limits. The $z'$ point in gray is not
  used for the model fit. The best-fit model spectrum is shown by the
  solid green line, and the corresponding magnitudes through the filters
  are indicated by asterisks.
  (\textsf{C}) The observed flux with the $\pm1\sigma$ uncertainty of
  the [O~{\sc iii}] line and the best-fit model prediction (asterisk).
  (\textsf{D}) The $3\sigma$ upper limit on the total infrared
  luminosity with a dust temperature of 40 K and an emissivity index
  of 1.5. Also shown is the best-fit model prediction (asterisk; zero IR
  luminosity due to absence of dust in the best-fit model).
  }
 \end{center}
\end{figure}

\clearpage

\begin{figure}
 \begin{center}
  \includegraphics[width=7cm]{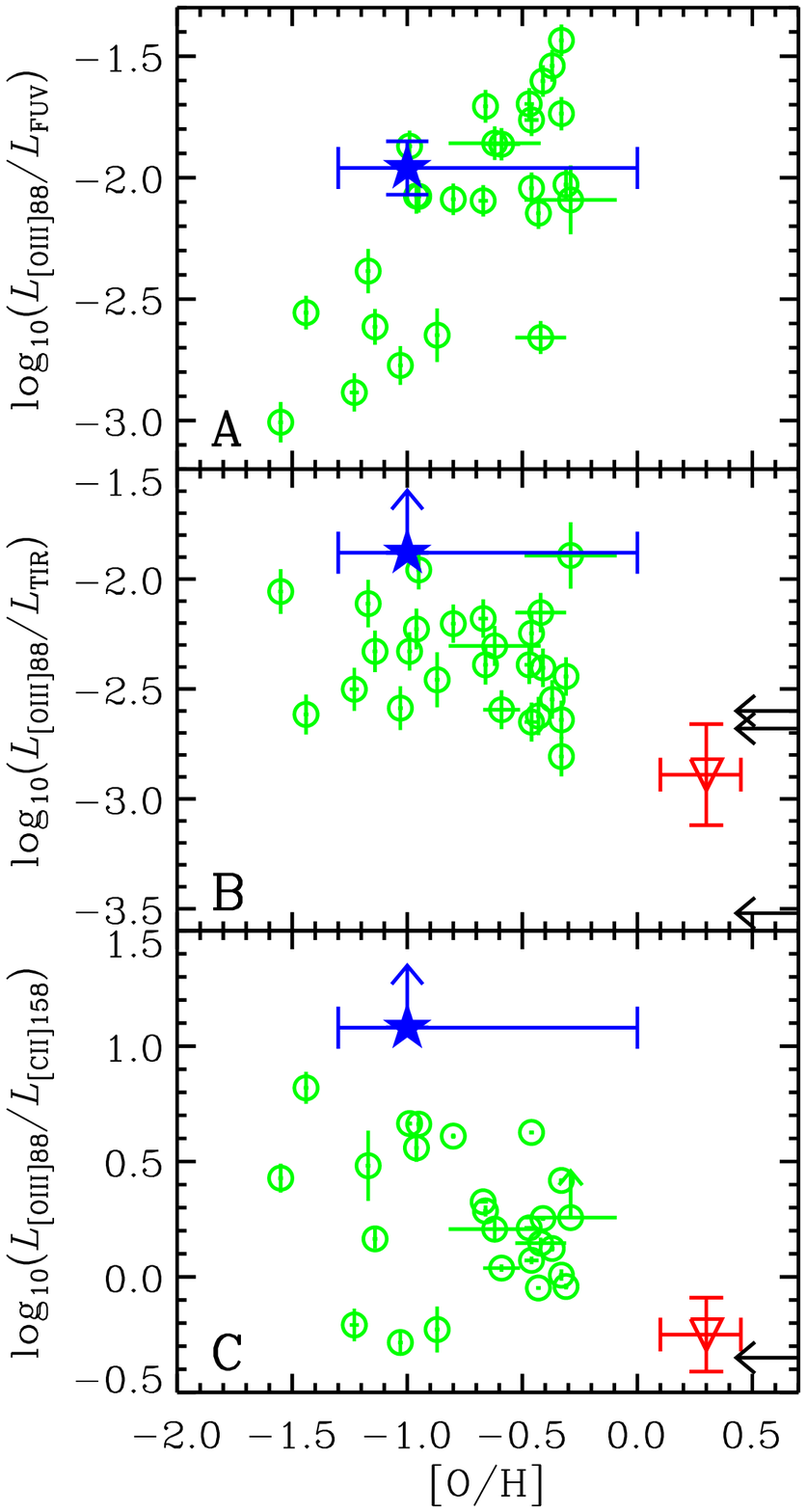}
  \caption{\textsf{Comparisons of SXDF-NB1006-2 and other galaxies
  detected in the [O~{\sc iii}] line.}
  The horizontal axis represents the oxygen abundance relative to the
  Sun on a logarithmic scale:  
  $[{\rm O/H}]=\log_{10}(n_{\rm O}/n_{\rm H})-\log_{10}(n_{\rm O}/n_{\rm
  H})_\odot$, where $n_{\rm O}$ and $n_{\rm H}$ are the number density
  of oxygen and hydrogen atoms, respectively, and the solar abundance is
  assumed to be 
  $12+\log_{10}(n_{\rm O}/n_{\rm H})_\odot=8.69$ \cite{asplund09}. 
  Circles with error bars represent data of nearby dwarf galaxies
  \cite{madden13,delooze14,cormier15}; inverted triangles with
  error bars are averages of nearby spiral galaxies \cite{brauher08}.
  The arrows at the right-side axis show luminosity ratios of dusty
  galaxies at $z\sim3$ to 4 whose oxygen abundances have not yet been
  measured \cite{ferkinhoff10,valtchanov11,delooze14}. Data from 
  SXDF-NB1006-2 are shown as five-pointed stars with error bars.
  (\textsf{A}) The [O~{\sc iii}]/far ultraviolet (FUV) luminosity
  ratio. The FUV luminosity is $\nu L_\nu$ at about 1500 \AA\ in the
  source rest-frame.
  (\textsf{B}) The [O~{\sc iii}]/total infrared (IR) luminosity
  ratio. The IR wavelength range is 8 to 1000 $\mu$m in the source
  rest-frame. Because the IR continuum of SXDF-NB1006-2 is not detected,
  we show a $3\sigma$ lower limit with a dust temperature of 40 K and an
  emissivity index of 1.5.
  (\textsf{C}) The [O~{\sc iii}]/[C~{\sc ii}] luminosity ratio. Because 
  the [C~{\sc ii}] 158 $\mu$m line of SXDF-NB1006-2 is not detected, we
  show a $3\sigma$ lower limit.
  }
 \end{center}
\end{figure}

\clearpage

 \begin{table}[h]
  \small
  \begin{center}
   \caption{\textsf{A summary of the observed and estimated properties
   of SXDF-NB1006-2.}}
   \begin{tabular}{lc}
    \hline \hline
    Right Ascension (J2000) &
    $2^\mathrm{h}18^\mathrm{m}56.536^\mathrm{s}$ 
    ($\pm 0.002^\mathrm{s}$) \\
    Declination (J2000) & $-5^{\circ}19'58.87''$ ($\pm 0.02''$) \\
    Redshift of [O~{\sc iii}] 88 $\mu$m & $7.2120 \pm 0.0003$ \\
    Ly$\alpha$ velocity shift (km s$^{-1}$) & $+1.1(\pm0.3)\times10^2$ \\
    $[$O~{\sc iii}$]$ 88 $\mu$m luminosity (W) $^*$ &
	$3.8(\pm0.8)\times10^{35}$ \\
    $[$C~{\sc ii}$]$ 158 $\mu$m luminosity (W) $^*$ & 
	$<3.2\times10^{34}$ ($3\sigma$) \\
    Total IR luminosity (W) $^*$ &
	$<2.9\times10^{37}$ ($3\sigma$) \\
    \hline
    Oxygen abundance $[\rm O/H]$ & $-1.0^{+1.0}_{-0.3}$ \\
    Star formation rate (log$_{10}$ M$_\odot$ yr$^{-1}$) $^\dag$ 
    & $2.54_{-0.71}^{+0.17}$ \\
    Star formation age (log$_{10}$ yr) & $6.00^{+1.00}$ \\
    Dust attenuation ($E_{B-V}$ mag) & $0.00^{+0.04}$ \\
    Escape fraction of ionizing photons & $0.54_{-0.54}^{+0.17}$ \\
    Stellar mass (log$_{10}$ M$_\odot$) $^\dag$ &
	$8.54_{-0.22}^{+0.79}$ \\
    \hline
   \end{tabular}
   \\
   $^{*}$ Assuming a concordance cosmology with $H_0=70$ km
   s$^{-1}$ Mpc$^{-1}$, $\Omega_{\rm M}=0.3$, and $\Omega_\Lambda=0.7$. \\
   $^{\dag}$ M$_\odot$ represents the Solar mass ($1.989\times10^{30}$ kg.)\\
  \end{center}
 \end{table}

\clearpage


\setcounter{page}{1}

\begin{center}
 {\LARGE Supplementary Materials for}

 \vspace{1cm}

 {\Large 
 {\bf Detection of an oxygen emission line from a high redshift galaxy
in the reionization epoch}
 }

 \vspace{1cm}

 Akio~K.~Inoue, Yoichi~Tamura, Hiroshi~Matsuo, Ken~Mawatari, 
 Ikkoh~Shimizu, Takatoshi~Shibuya, Kazuaki~Ota, Naoki~Yoshida, 
 Erik~Zackrisson, Nobunari~Kashikawa, Kotaro~Kohno, 
 Hideki~Umehata, Bunyo~Hatsukade, Masanori~Iye, Yuichi~Matsuda, 
 Takashi~Okamoto, Yuki~Yamaguchi

 \vspace{1cm}

 Correspondence to: akinoue@las.osaka-sandai.ac.jp
\end{center}

\vspace{1cm}

\noindent
{\bf The PDF file includes:}

Materials and Methods

Figs. S1 to S12

Tables S1 to S4

\clearpage

\section*{Materials and Methods}

\section{ALMA observation and results}

\subsection{Observations and data reduction}

The ALMA band 8 data of the [O~\textsc{iii}] 88 $\mu$m emission line
(rest-frame frequency 3393.01~GHz) redshifted to 413~GHz for the
target galaxy, SXDF-NB1006-2, were obtained on 2015 June 7, 9 and 14 
(cycle 2, project ID: 2013.1.01010.S, PI: A.~K.\ Inoue). Thirty-seven
to 41 operational antennas were employed with the C34-6/7 array
configuration, where the maximum and minimum baseline lengths were
783.5~m and 21.3~m, respectively. The correlator was configured so that
400.1--403.6 and 412.1--414.0 GHz were covered by four spectral windows,
each of which was used in the Frequency Division Mode (FDM) with a
1.875~GHz bandwidth and a 7.8125~MHz (5.67 km~s$^{-1}$ at 413~GHz)
resolution. A total of 2.0 hour was spent for on-source integration
under excellent atmospheric conditions with precipitable water vapors
(PWVs) of 0.4--0.5 mm. The resulting spatial resolution with the natural
weighting is $0.''35 \times 0.''26$ (in full width at half maximum
(FWHM); position angle PA = $+82^{\circ}$), with the r.m.s.\ noise
levels of 0.53 and 0.042 mJy~beam$^{-1}$, respectively, for the 
20 km~s$^{-1}$ resolution cube and the continuum image. Two quasars,
J0241$-$0815 ($S_\mathrm{413\,GHz}=1.6$~Jy, 6$^{\circ}$ away from the
target) and J2232+1143 (0.3~Jy), and Ceres were used for complex gain,
bandpass and flux calibration, respectively. The flux calibration
accuracy is estimated at 10\%.

The band 6 data targeting the [C~\textsc{ii}] 158 $\mu$m line (the
rest-frame frequency of 1900.54~GHz) at 231 GHz for SXDF-NB1006-2 were
obtained on 2014 August 1 and 5 (cycle 1, project ID:
2012.1.00374.S, PI: K.\ Ota), where 30--34 antennas were operational
under the C32-5 configuration (the maximum and minimum baseline lengths
of 558.2~m and 17.2~m, respectively). The correlator was configured to
cover 215.7--219.5 and 230.4--234.2 GHz in the FDM 1.875~GHz mode with a
0.488 MHz (0.63 km~s$^{-1}$ at 231~GHz) resolution. The conditions were
reasonable (PWV = 1--2 mm) during the on-source time of 1.8 hour. The
resulting synthesized beam size (FWHM) with the natural weighting is
$0.''80 \times 0.''60$ (PA = $-81^{\circ}$). The achieved noise levels
for the 20 km~s$^{-1}$ cube and the continuum image are 0.26 and 0.014
mJy~beam$^{-1}$, respectively. Complex gain calibration was made using a
nearby quasar J0215$-$0222 ($S_\mathrm{231~GHz}=0.06$~Jy, $4^{\circ}$
away from the target), while three quasars (J0006$-$0623, J0423$-$0120
and J0241$-$0815) were used for bandpass calibration. Both Neptune and
J0238+166 were used for flux calibration to cross-check the amplitude
scaling. The flux calibration accuracy is estimated at 8\%.

We calibrated the raw visibility data in a standard manner using the
CASA software \cite{mcmullin07} version 4.3.1 and 4.2.1 for the
[O~\textsc{iii}] and [C~\textsc{ii}] data, respectively, along with
a standard calibration script provided by the observatory.  In addition
to standard flagging such as shadowed antennas, manual flagging has
carefully been made for low-gain antennas and abnormal visibilities.
For the [O~\textsc{iii}] (band 8) data, Earth's atmospheric ozone lines
severely affect up to 10\% of the frequency coverage in 3 out of 4
spectral windows and are flagged properly, while the rest of the
spectral window where the [O~\textsc{iii}] line is expected does not
suffer from the atmospheric contamination and remains unflagged.

Imaging is carried out using a CASA task, \texttt{clean}, with the
natural weighting to maximize the point-source sensitivities.  Continua
are not subtracted in [O~{\sc iii}] and [C~{\sc ii}] imaging because no
continuum emission is found.  As the [O~\textsc{iii}] emission is found
to be marginally resolved with the naturally-weighted beam (the
intrinsic source size from a Gaussian fit of $0.''4 \times 0.''3$, PA
$\simeq 90^{\circ}$), we also make a $uv$-tapered image with
\texttt{outertaper} = $0.''3$ to achieve a good detection.  The
resulting beam size is $0.''45 \times 0.''38$ (PA = $+78^{\circ}$).
Synthesized-beam deconvolution was made for the [O~\textsc{iii}] image
using the CLEAN algorithm down to a $1.5\sigma$ level.

\subsection{Results}

\subsubsection{[O~{\sc iii}] 88 $\mu$m line}

The [O~\textsc{iii}] emission is detected at a significance of
$5.3\sigma$ at the position where the Ly$\alpha$ emission is detected
(Figure~1A). The $uv$-tapered image is integrated over $-300$ to $+230$
km s$^{-1}$ with respect to the [O~\textsc{iii}] redshift of
$z_\mathrm{[OIII]} = 7.2120$,  which is obtained by a Gaussian fit
to the spectrum of a velocity resolution of 20 km s$^{-1}$. The
histogram of pixel signal-to-noise ratios (SNRs) in the [O~\textsc{iii}]
integrated intensity image (Figure~S1) is well described by a Gaussian
(i.e., normal distribution) at the pixel values below SNR $<4$, while
the number of pixels with positive fluxes surpasses that of the negative
pixels at SNR $>4$. This is due to the contribution from the real
[O~\textsc{iii}] emission line. To further test the significance of the
detection, we separately image the data taken during three independent
tracks made on 2015 June 7, 9 and 14. The on-source time of each track
is 40 min. We find a $\sim 3\sigma$ peak at the position of
SXDF-NB1006-2 in every image, demonstrating a robust detection of the
[O~\textsc{iii}] line (Figure~S2).

\begin{figure}[htbp]
\begin{center}
 \includegraphics[width =
 0.4\textwidth,keepaspectratio,clip,angle=-90]{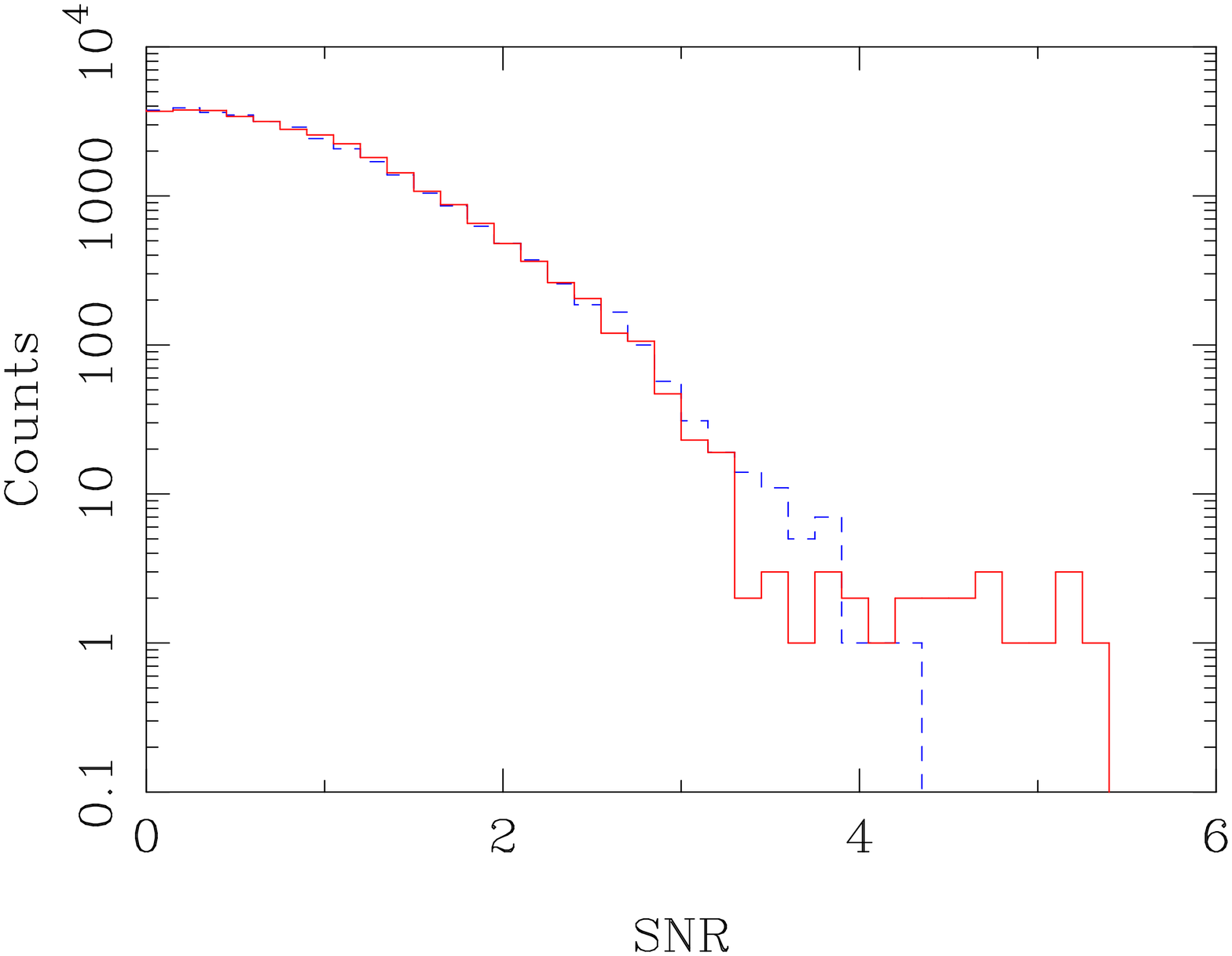}
 \vspace{0.5cm}\\
  \begin{minipage}{14cm}
   Figure~S1: \textsf{Histogram of pixel signal-to-noise ratios (SNRs)
   of the [O~{\sc iii}] integrated intensity map.} The data are taken
   from the entire field of view of ALMA band 8 observations. Positive
   flux values are shown by the red solid line, while negative values
   are shown by the blue dashed line. The histograms are well described
   by a Gaussian up to SNRs around 4, whereas there is an excess in
   positive flux values at SNR $>4$, to which the [O~\textsc{iii}] 
   emission contributes.
  \end{minipage}
\label{fig-016_fluxhist}
\end{center}
\end{figure}

\begin{figure}[htbp]
\begin{center}
\includegraphics[width =
 0.4\textwidth,keepaspectratio,clip,angle=-90]{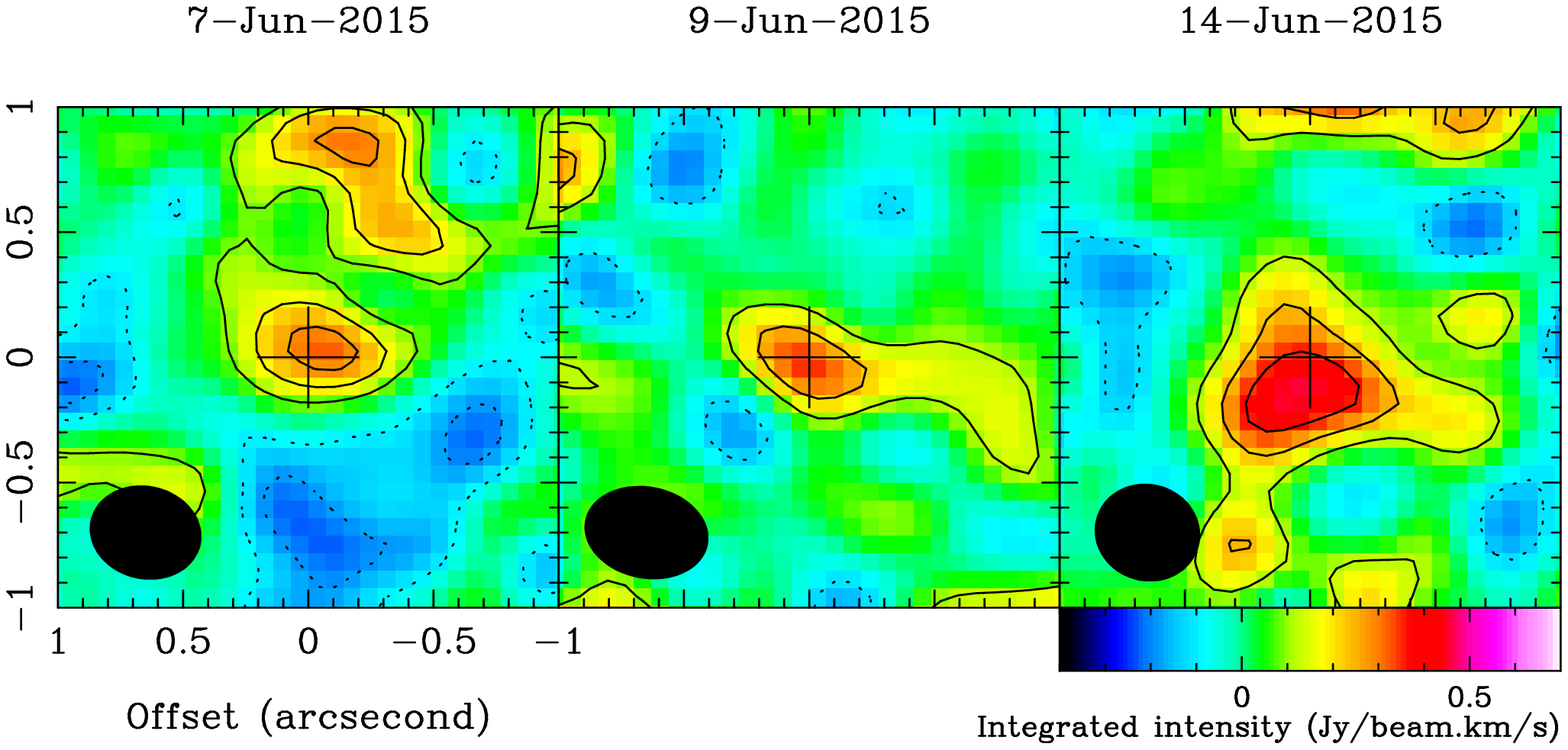}
 \vspace{0.5cm}\\
  \begin{minipage}{14cm}
   Figure~S2: \textsf{[O~\textsc{iii}] 88 $\mu$m integrated intensity
   maps of each observing date.} Every image shows a $\sim 3\sigma$
   peak at the position where the [O~\textsc{iii}] line is found (crosses),
   demonstrating the detection robustness. Contours start from $1\sigma$
   with a step of $1\sigma$. The dotted contours show negative values. The
   ellipse at the bottom-left corner on each panel indicates the ALMA beam
   size.
  \end{minipage}
\label{fig-008_2pr}
\end{center}
\end{figure}

\begin{figure}[htbp]
\begin{center}
\includegraphics[width =
 0.6\textwidth,keepaspectratio,clip]{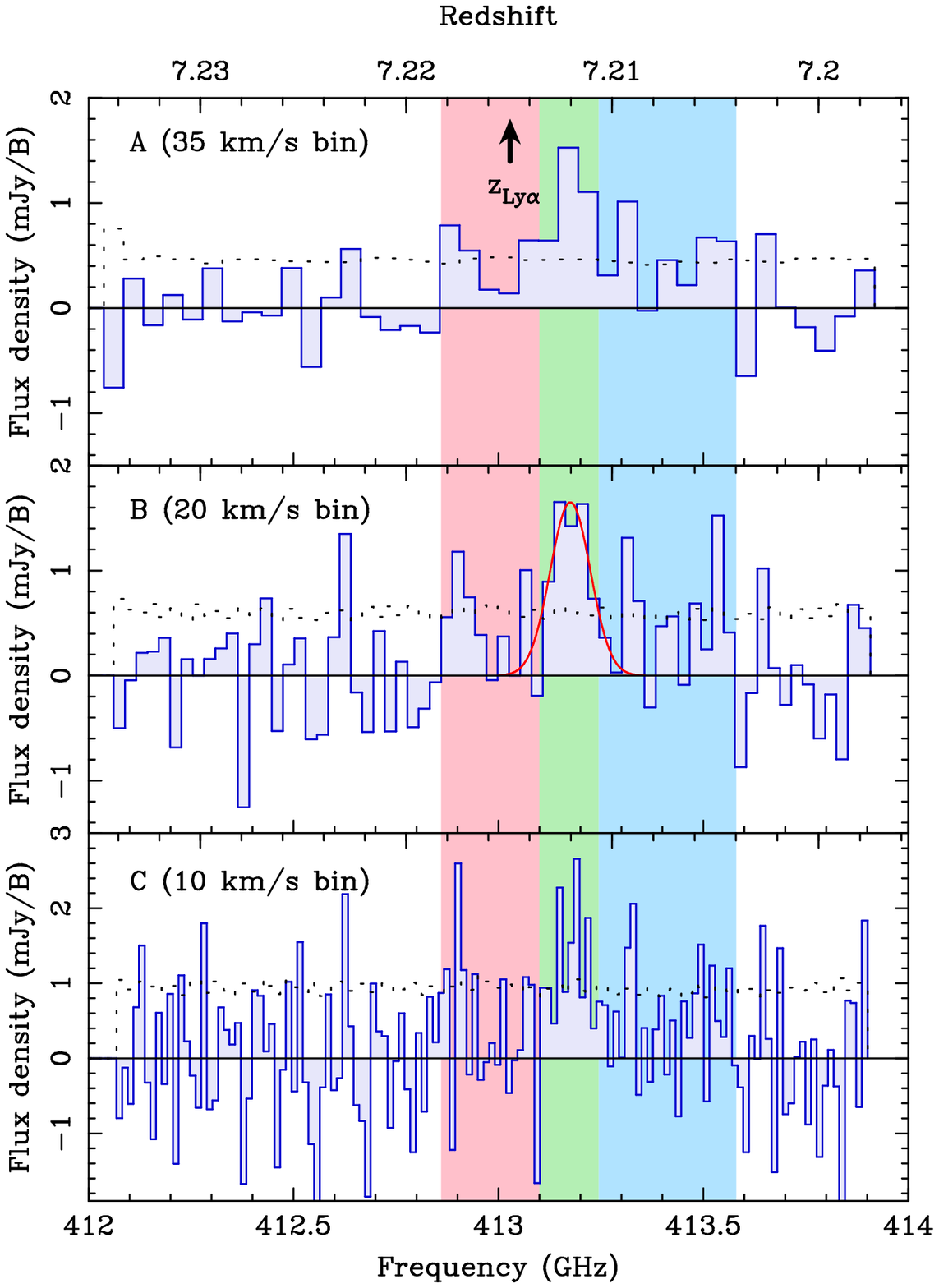}
 \vspace{0.5cm}\\
  \begin{minipage}{14cm}
   Figure~S3: \textsf{[O~\textsc{iii}] 88 $\mu$m spectra at the
   intensity peak with different velocity resolutions.} From A to
   C, the velocity resolution is 35, 20 and 10 km~s$^{-1}$. The
   dotted line is the r.m.s.\ noise level measured over each resolution
   element. In panel A, the Ly$\alpha$ redshift is indicated with the
   upward arrow. The vertical bands in red, green and blue represent the
   velocity intervals over which the integrated intensity images shown
   in Figure~S5 are integrated. In the panel B, we show the
   best-fit Gaussian profile for the emission line. The FWHM is
   $\approx80$ km s$^{-1}$.
  \end{minipage}
\label{OIII.tap0.3arcsec_x66y65_v2}
\end{center}
\end{figure}

In the band 8 spectra at the intensity peak position, a narrow (FWHM of
$\approx80$ km~s$^{-1}$) line feature is evident at around 413.2 GHz
(Figure~S3B). This line feature is neither a collection of
spurious spikes nor a part of spectral baseline wiggles 
(Figure~S3C). The redshift of the line is measured as 
$z_{\rm [OIII]}=7.2120\pm0.0003$, slightly lower than the redshift
determined from the Ly$\alpha$ emission line \cite{shibuya12}. This
redshift difference corresponds to a velocity offset of $\approx 110$
km~s$^{-1}$ (\S2.1), which is reasonably accounted for when the bluer
(i.e., shorter-wavelength) part of the Ly$\alpha$ emission line is
attenuated by the IGM along the sightline, as reported for many
Ly$\alpha$ emitters at $z \sim 6$--7 \cite{kashikawa06,shibuya12}, 
in addition to the ISM attenuation \cite{hashimoto13,erb14,shibuya14b}.

We measure the total flux density of the [O~\textsc{iii}] emission by 
fitting the tapered integrated intensity image to a Gaussian using a 
CASA task, \texttt{imfit}, and deconvolving the clean beam to derive 
the intrinsic source flux.  Table~S1 lists the integrated
intensity ($0.45 \pm 0.09$ Jy~km~s$^{-1}$, which corresponds to a flux
of $6.2 \times 10^{-21}$ W~m$^{-2}$) and luminosity 
($9.8\times 10^{8}~L_{\odot}$), where $L_\odot=3.8\times10^{26}$ W
is the solar luminosity. The [O~\textsc{iii}] line luminosity is 
at the high end of the detections made in local dwarf galaxies
\cite{cormier15}, normal spirals \cite{brauher08} and (ultra-)luminous
infrared galaxies \cite{gracia-carpio11}, while it is an order of
magnitude lower than (demagnified) [O~\textsc{iii}] line luminosities
found in gravitationally-lensed dusty starburst galaxies at $3 < z < 4$
\cite{ferkinhoff10,valtchanov11}.

\begin{table}[t]
\label{table:alma_line}
\begin{center}
 Table~S1: \textsf{ALMA results of [O~{\sc iii}] 88 $\mu$m and [C~{\sc
 ii}] 158 $\mu$m lines of SXDF-NB1006-2.}\\
\begin{tabular}{ccc} 
\hline \hline
&[O~\textsc{iii}] 88 $\mu$m  &[C~\textsc{ii}] 158 $\mu$m \\
\hline
Integrated intensity (Jy km s$^{-1}$) & $0.45 \pm 0.09$ & $< 0.069$
	 $(3\sigma)$ \\
Flux calibration uncertainty & $10\%$ & $8\%$ \\
Flux (W m$^{-2}$)       & $(6.2 \pm 1.4) \times 10^{-21}$ $^\dag$ & $< 5.3
	 \times 10^{-22}$ $(3\sigma)$ \\
Luminosity ($L_{\odot}$) $^*$  & $(9.8 \pm 2.2) \times 10^{8}$ $^\dag$
     & $< 8.3 \times 10^{7}$ $(3\sigma)$\\
 Beam-deconvolved source size & $0.4'' \times 0.3''$ (PA $\simeq
     90^{\circ}$) & --- \\
\hline
\end{tabular}
\\
$^*$ Assuming a concordance cosmology with $H_0=70$ km s$^{-1}$
 Mpc$^{-1}$, $\Omega_{\rm M}=0.3$, and $\Omega_\Lambda=0.7$. \\
$^\dag$ Flux calibration uncertainty is included in the error.
\end{center}
\vspace{-0.5cm}
\end{table}

\subsubsection{[C~{\sc ii}] 158 $\mu$m line}

In the integrated intensity map of the [C~\textsc{ii}] emission summed
over the same velocity range as that of the [O~\textsc{iii}] image, we
find no [C~\textsc{ii}] emission with a $>3\sigma$ significance around
SXDF-NB1006-2 (Figure~S4A). Thus, we conclude that
there is no significant [C~{\sc ii}] line source integrated over the
same velocity range as the [O~{\sc iii}] line at the position
emitting the [O~{\sc iii}] and hydrogen Ly$\alpha$ lines. Thus, we 
place a $3\sigma$ upper limit on flux and luminosity for the 
[C~{\sc ii}] line (Table~S1). On the other hand, we notice that when
the band 6 cube is integrated over two velocity ranges, $-20 < v < 260$
and $90 < v < 230$ km~s$^{-1}$, low-significance ($3.5\sigma$ and
$3.7\sigma$) bumps appear close to the LAE (denoted as `NE' and `SE' in
Figure~S4B and S4C, respectively). Unfortunately, the
features are severely affected by the Earth's atmospheric ozone line at
231.28 GHz, which prevents us from judging whether or not these
are spurious. Furthermore, there are a few more $3\sigma$--$4\sigma$
enhancements remaining over the map (see a $3.9\sigma$ enhancement at
the northern edge of Figure~S4A). 

\begin{figure}[htbp]
\begin{center}
\includegraphics[width =
 0.4\textwidth,keepaspectratio,clip,angle=-90]{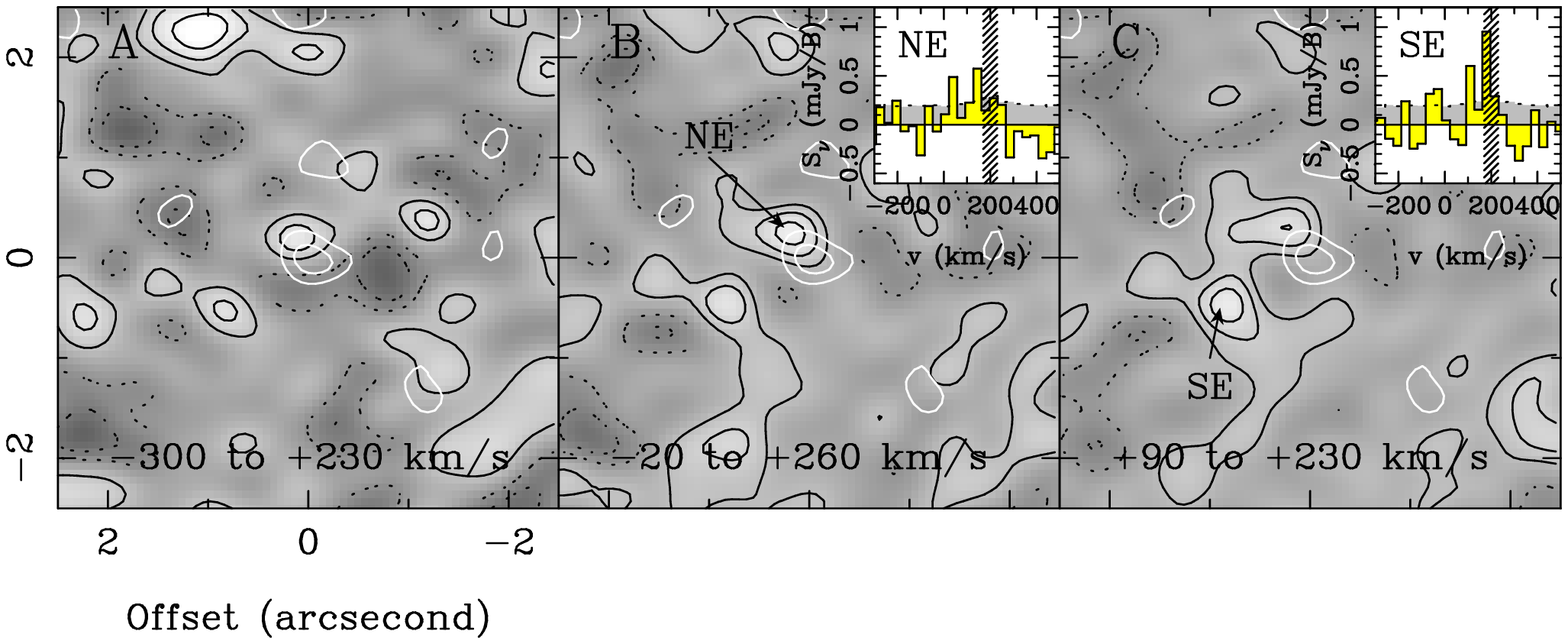}
 \vspace{0.5cm}\\
  \begin{minipage}{14cm}
   Figure~S4: \textsf{[C~{\sc ii}] 158 $\mu$m emission line integrated
   intensity maps of SXDF-NB1006-2.} (A) The integrated intensity image
   of the [C~{\sc ii}] line over the same velocity range as that of the
   [O~{\sc iii}] line ($-300<v<+230$ km~s$^{-1}$). The black contours
   represent ($\pm 1$, $\pm 2$, $\pm 3$, ...)$\times \sigma$, where
   $\sigma =$ 28 mJy~beam$^{-1}$~km~s$^{-1}$. Negative contours are
   shown by dotted lines. The white contours show the [O~{\sc iii}] line
   image and are drawn at $2\sigma$ and $4\sigma$. No significant
   [C~{\sc ii}] emission is found.
   (B) The same as A but integrated over $-20< v<+260$ km~s$^{-1}$
   ($\sigma =$ 23 mJy beam$^{-1}$ km s$^{-1}$). The inset shows the
   spectrum at the marginal $3.5\sigma$ enhancement north-east to the
   [O~{\sc iii}] position (denoted as `NE'). The dotted line with gray
   shade shows the $1\sigma$ noise level.  The frequency range where an
   atmospheric absorption line contaminates the spectrum is indicated by a
   hatched band. 
   (C) The same as A but integrated over $+90<v<+260$  km~s$^{-1}$
   ($\sigma =$ 21 mJy~beam$^{-1}$~km~s$^{-1}$). The inset shows the
   spectrum at the marginal $3.7\sigma$ enhancement south-east to the
   [O~{\sc iii}] position (denoted as `SE').
  \end{minipage}
\label{cii}
\end{center}
\end{figure}

\subsubsection{Dust continuum}

Continuum emission remains undetected in both the band 8 (735 $\mu$m)
and band 6 (1.33 mm) images. The $3\sigma$ upper limits measured for
naturally-weighted images are 0.12 and 0.042 mJy at 735 $\mu$m and 1.33
mm, respectively. The total IR luminosity assuming a modified blackbody
\cite{debreuck03} integrated over the rest-frame wavelengths of 8--1000
$\mu$m is estimated to be $L_{\rm TIR}<1\times10^{11}~L_\odot$, where
the dust temperature and emissivity index are assumed to be $T_{\rm
dust}=40$ K and $\beta = 1.5$, respectively. This limit can be relaxed
to $<2\times10^{11}~L_{\odot}$ for a higher dust temperature of
$T_\mathrm{dust} = 50$ K, while if the galaxy has cooler dust
($T_\mathrm{dust} = 30$ K), the luminosity limit obtained from the 1.33
mm photometry becomes more stringent ($<4\times 10^{10}~L_{\odot}$). 
Table~S2 is a summary of these results.
The emissivity index of 1.5 which we assumed is a typical value
observed in nearby star-forming galaxies \cite{dunne01}. It is reported
that the typical star-forming galaxies ($L\sim L_*$) at $z\sim4$ have 
$T_{\rm dust}\simeq30$ K \cite{lee12}. A bright LAE at $z\simeq7$,
Himiko \cite{ouchi09}, is estimated to have $T_{\rm dust}=30$--40 K
\cite{hirashita14}. We therefore assume $T_{\rm dust}=40$ K as a
fiducial value for SXDF-NB1006-2 in this paper. For this temperature,
the effect of the cosmic microwave background whose temperature is
22 K at $z=7.2$ is small \cite{dacunha13}.

\begin{table}[t]
\label{table:alma_continuum}
\begin{center}
 Table~S2: \textsf{ALMA results of the dust IR continuum of SXDF-NB1006-2.}\\ 
\begin{tabular}{ccc} 
\hline \hline
& Band 8 (735 $\mu$m)  & Band 6 (1.33 mm) \\
\hline
Flux density (mJy) & $< 0.12$ ($3\sigma$) & $< 0.042$ ($3\sigma$) \\
\hline
Dust temperature (K) & \multicolumn{2}{c}{Total infrared luminosity
     ($L_\odot$, $3\sigma$) $^*$} \\
\hline
30  & $< 9.0\times 10^{10}$  & $< 3.8\times 10^{10}$ \\
40  & $< 1.1\times 10^{11}$  & $< 8.3\times 10^{10}$ \\
50  & $< 1.7\times 10^{11}$  & $< 1.7\times 10^{11}$ \\
\hline
\end{tabular}
\\
$^*$ We assume a single-temperature modified blackbody with an
 emissivity index of 1.5 and a concordance cosmology with $H_0=70$ km
 s$^{-1}$ Mpc$^{-1}$, $\Omega_{\rm M}=0.3$, and $\Omega_\Lambda=0.7$.
\end{center}
\end{table}

\subsubsection{Possible kinematics signature in the [O~{\sc iii}] spectrum}

The [O~\textsc{iii}] image (Figure~1A) is marginally resolved and
likely elongated in the east--west direction. The beam-deconvolved
source size, if the source is approximated by a two-dimensional
Gaussian, is estimated to be $0.''4 \times 0.''3$ in FWHM (PA
$\sim90^{\circ}$), corresponding  to a physical scale of $\simeq 2
\times 1.5$ kpc$^2$. This extended structure may be attributed to
high-velocity components that are seen as red-shifted and
blue-shifted marginal broad signals in the [O~{\sc iii}] spectrum
(Figure~S3A). We made images of the central narrow component
($-50 < v < +50$ km~s$^{-1}$) and the red-shifted and blue-shifted 
marginal high-velocity components ($50<v<230$ km~s$^{-1}$ and
$-300<v<-50$ km~s$^{-1}$; Figure~S5). Although the SNR is not high
enough, it seems that the high-velocity components are mostly overlapped
but exhibit a small spatial offset of $0.''3$ ($\simeq1.5$ kpc), which
is larger than the statistically-expected positional uncertainty
($\simeq 0.5\theta/\mathrm{SNR} \simeq 0.05''$, where $\theta$ is the
beam size).

\begin{figure}[htbp]
\begin{center}
\includegraphics[width =
 0.5\textwidth,keepaspectratio,clip,angle=-90]{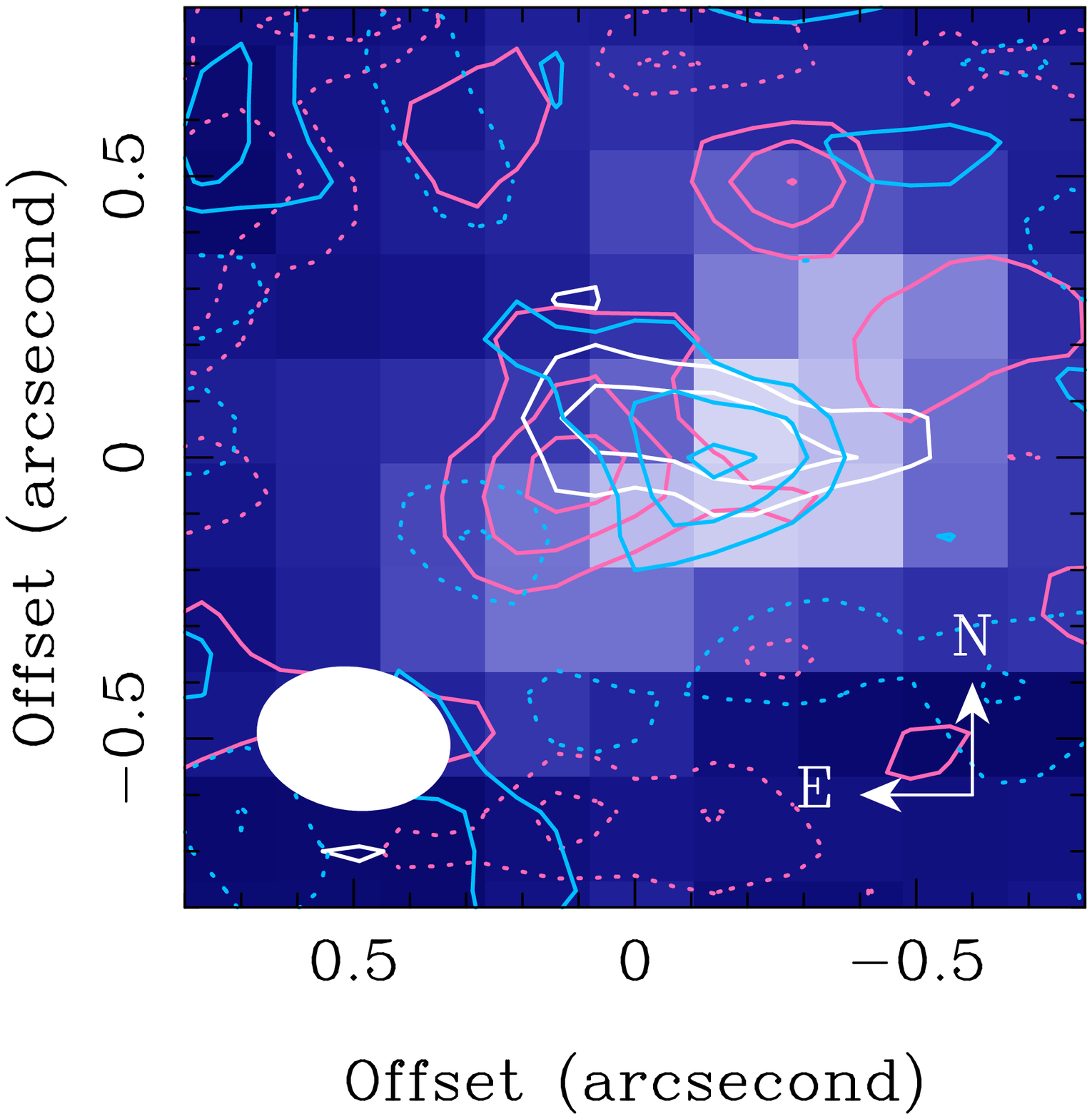}
 \vspace{0.5cm}\\
  \begin{minipage}{14cm}
   Figure~S5: \textsf{Spatial distribution of different velocity
   components of the [O~\textsc{iii}] 88 $\mu$m emission.} Red and blue
   contours show the red-shifted ($50 < v < 230$ km~s$^{-1}$) and
   blue-shifted ($-300 < v < -50$ km~s$^{-1}$) marginal components,
   respectively, while white contours represent the narrow component 
   ($-50 < v < +50$ km~s$^{-1}$) which likely traces the systemic
   redshift of SXDF-NB1006-2. These velocities are measured with respect
   to the [O~\textsc{iii}] line peak at $z = 7.2120$. Contours are drawn
   at ($\pm1$, $\pm2$, $\pm3$, ...)$\times \sigma$ for the high-velocity
   components and ($\pm2$, $\pm3$)$\times \sigma$ for the narrow
   component for clarity. Negative contours are shown by the dotted
   lines. The background image is the Subaru narrowband Ly$\alpha$ image
   \cite{shibuya12}. The ellipse at the bottom-left corner represents
   the naturally-weighted beam size of ALMA.
  \end{minipage}
\label{fig-017_kin}
\end{center}
\end{figure}

A possible explanation of these marginal high-velocity component is
rotating motion of gravitationally-bounded gas, which is often observed
in high-$z$ massive galaxies \cite{wang13,venemans15}. A dynamical mass
is an order of   
$M_{\rm dyn}\sim1\times10^5~v_{\rm circ}^2 D$ M$_{\odot}$, where 
$v_{\rm circ} = 0.75~\Delta v~(\sin{i})^{-1}$ is the circular velocity
in units of km~s$^{-1}$ ($\Delta v$ and $i$ are, respectively, the line
FWHM and the inclination angle) and $D$ is the diameter of the galaxy
measured in kpc. If the possible red/blue-shifted components of
the [O~\textsc{iii}] line is produced by a rotating disk with a
$D\simeq2$ kpc and a velocity width of $\Delta v=400$ km~s$^{-1}$
(FWHM), the dynamical mass is estimated to be 
$M_{\rm dyn}\sim5\times10^{10}~M_{\odot}$, where we assume that 
the galaxy is a circular disk and the intrinsic source size gives the
inclination angle, i.e., $i = \cos^{-1}{(0.''3/0.''4)}$. The dynamical
mass is 2 orders of magnitude larger than the best-fit stellar mass
obtained from the SED fitting (\S3.5). However, it can not be excluded
that a passive stellar population of $<5\times10^{10}$ M$_\odot$
coexists with a young starburst in SXDF-NB1006-2 (Figure~S11). In
a cosmological simulation \cite{shimizu14,shimizu15}, galaxies at
$z=7.2$ with similar UV luminosities to SXDF-NB1006-2 have a stellar mass
of a few $\times10^{10}$ M$_\odot$ (see \S4, Figure~S10) and a dark
halo mass of a few $\times10^{11}$ M$_{\odot}$. Therefore, the
high-velocity component could be explained by rotational motion.

Another interpretation is violent motions such as outflows driven by
stellar winds and supernovae in the star forming region of
SXDF-NB1006-2, which is probably seen as the narrow component  
(FWHM of $\approx80$ km s$^{-1}$). Yet another possibility is turbulent
motion driven by, for example, a merger event which might mimic the
spatial offset, although we find a single unresolved component in
the $J$ band image tracing the rest-frame UV continuum (Figure~2A). 
However, we can not conclude the origin of the possible
high-velocity components because of the limited SNR for the moment.

\section{Optical-to-near infrared data}

The target galaxy, SXDF-NB1006-2, is in the Subaru/XMM-Newton Deep
Survey Field (SXDF) \cite{furusawa08} where multi-wavelength deep
observations have been carried out. We have gathered archival
optical-to-near infrared (NIR) deep images available in the SXDF:
Subaru/Suprime-Cam broadband $z'$ \cite{furusawa16} and narrowband
$NB1006$ \cite{shibuya12}, UKIRT/WFCAM broadband $J$, $H$, and $K$ taken
in the UKIRT Infrared Deep Sky Survey (UKIDSS) Ultra-Deep Survey (UDS)
\cite{lawrence07}, and {\it Spitzer}/IRAC $3.6~\mu$m and $4.5~\mu$m
taken in the {\it Spitzer} Extended Deep Survey (SEDS)
\cite{ashby13}. We measured the point spread functions (PSFs) using
stellar objects in these images, resulting in FWHMs of $1.''0$ ($z'$),
$0.''4$ ($NB1006$), $0.''8$ ($J$, $H$, and $K$), and $1.''8$ ($3.6\mu$m
and $4.5\mu$m). Except for the $NB1006$ image, the photometry was
performed using $2\times$PSF (FWHM) apertures because the object is
almost unresolved or not detected. For the $NB1006$ image, we performed
Kron photometry \cite{kron80} with the parameter $k=2$ to obtain a total
flux density from the spatially extended Ly$\alpha$ emission. The
photometric measurements are summarized in Table~S3. The magnitudes are
the AB system \cite{oke90}. 

\begin{table}[htb]
 \begin{center}
  Table~S3: \textsf{Photometric data of SXDF-NB1006-2.}
 \begin{tabular}{cccc} 
  \hline \hline
  Band & Wavelength ($\mu$m) & PSF FWHM ($''$) & Magnitude (AB) \\
  \hline
  $z'$ & 0.91 & $1.''0$ & $>27.06$ $^*$ \\
  $NB1006$ & 1.00 & $0.''4$ & $24.50\pm0.22$ $^\dag$ \\
  $J$ & 1.26 & $0.''8$ & $25.46\pm0.18$ $^\ddag$\\
  $H$ & 1.65 & $0.''8$ & $>25.64$ $^*$\\
  $K$ & 2.23 & $0.''8$ & $>25.84$ $^*$\\
  $IRAC3.6$ & 3.54 & $1.''8$ & $>24.64$ $^*$\\
  $IRAC4.5$ & 4.49 & $1.''8$ & $>24.26$ $^*$\\
  \hline
 \end{tabular}
  \\
  $^*$ 3$\sigma$ lower limit in $2\times$PSF circular aperture.\\
  $^\dag$ Kron magnitude with a $2.''64\times1.''32$ ellipse aperture.\\
  $^\ddag$ $2\times$PSF circular aperture.\\
  \end{center}
\end{table}

\subsection{Velocity offset between Ly$\alpha$ and [O~{\sc iii}] lines}

We have performed a profile fitting of the Ly$\alpha$ line with an
asymmetric Gaussian function \cite{shibuya14b}: 
\begin{equation}
 F_\lambda = A\exp\left[
  \frac{-(\lambda-\lambda_0)^2}{2\{\sigma+a(\lambda-\lambda_0)\}^2}
 \right]\,,
\end{equation}
where $A$ is the peak flux, $\lambda_0$ is the peak wavelength, $\sigma$
is the line width, and $a$ is the asymmetric parameter. If $a>0$, the
blue part of the line profile is weakened as the Ly$\alpha$ line observed
in high-$z$ \cite{kashikawa06}. The usual Gaussian function is recovered
with $a=0$. First, we have made a fitting with a Gaussian function and
obtained the following results: $A=(1.67\pm0.15)\times10^{-18}$ erg
s$^{-1}$ cm$^{-2}$ \AA$^{-1}$, $\lambda_0=9987.51\pm0.822$ \AA, and 
$\sigma=4.71\pm0.42$ \AA\ (Figure~S6). The corresponding redshift is 
$z_{\rm Ly\alpha}=7.2156\pm0.0007$ (Gaussian fit). Next, we have made
a fitting with an asymmetric Gaussian function with a fixed $\sigma=4.71$
\AA\ from the Gaussian fit, which is also consistent with the observed
FWHM of the line (11.5 \AA\ \cite{shibuya12}). The results are
$A=(1.65\pm0.16)\times10^{-18}$ erg s$^{-1}$ cm$^{-2}$ \AA$^{-1}$,
$\lambda_0=9986.67\pm0.967$ \AA, and $a=0.169\pm0.065$ (Figure~S6). The
corresponding redshift is $z_{\rm Ly\alpha}=7.2150\pm0.0008$
(asymmetric Gaussian fit). Assuming that the [O~{\sc iii}] 88 $\mu$m
line at $z_{\rm [OIII]}=7.2120\pm0.0003$ traces the systemic redshift, 
we have obtained the velocity offset of the Ly$\alpha$ line 
$\Delta v_{\rm Ly\alpha}=+(1.1\pm0.3)\times10^2$ km s$^{-1}$, where we
have corrected the Ly$\alpha$ redshift for the heliocentric motion of
Earth at the observing date ($+4$ km s$^{-1}$). Note that the ALMA
spectrum of the [O~{\sc iii}] line is already corrected for the Earth's
motion in the data reduction process.

\begin{figure}
 \begin{center}
  \includegraphics[width=8cm]{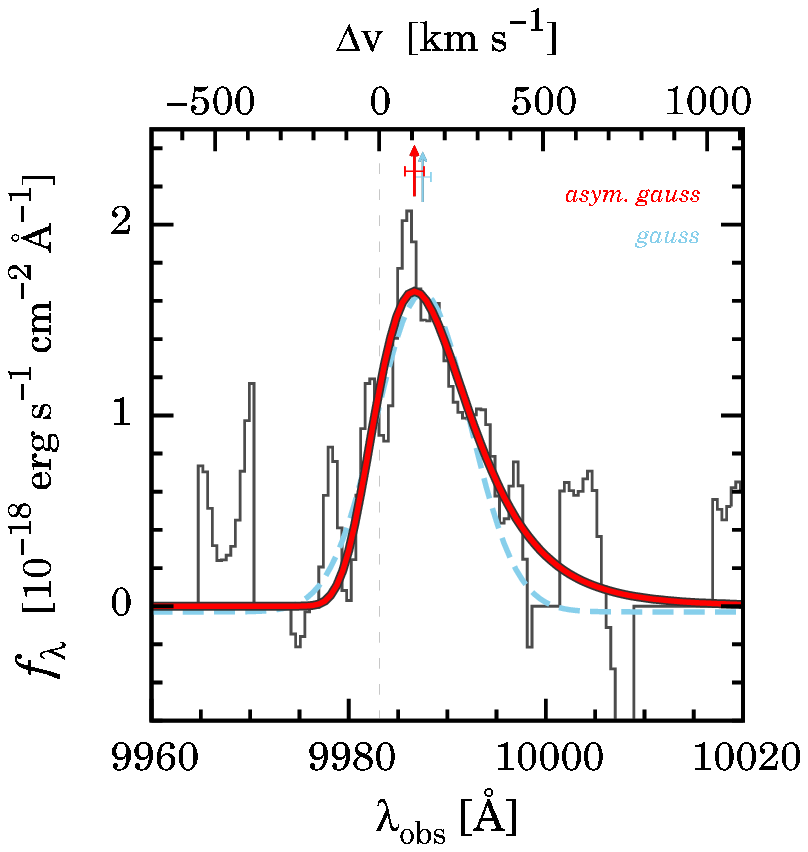}
  \vspace{0.5cm}\\
  \begin{minipage}{14cm}
   Figure~S6: \textsf{Ly$\alpha$ line profile fitting results.} 
   The black solid line is the observed spectrum and the red solid
   line is the best-fit profile with an asymmetric Gaussian function
   expressed in equation (1). The cyan dashed line is the best-fit
   result with a simple Gaussian function for a reference. The central
   wavelengths with their $\pm1\sigma$ uncertainties are shown by the 
   upward arrows with error-bars. The wavelength range used in the
   fitting is 9975--9998 \AA. The top horizontal axis is the velocity
   shift relative to the systemic redshift $z=7.2120$ measured from the 
   [O~{\sc iii}] 88 $\mu$m line and corrected for the heliocentric
   motion of Earth.
  \end{minipage}
 \end{center}
\end{figure}

\subsection{Empirical SFR estimation}

We now estimate the SFR of SXDF-NB1006-2 with empirical relations.
We assume a Salpeter initial mass function (IMF) \cite{salpeter55} with
the mass range of 0.1--100 M$_\odot$ throughout this paper. There
is a good correlation between the [O~{\sc iii}] 88 $\mu$m line
luminosity and the SFR derived from the sum of the FUV and IR
luminosities based on a large compilation of various kinds of galaxies
including nearby low-metallicity dwarfs, ULIRGs, AGNs, and high-$z$
dusty starbursts \cite{delooze14}. The [O~{\sc iii}]--SFR relations
for specific kinds of galaxies are slightly different from each other. 
If we assume the relation derived from the entire sample of
\cite{delooze14}, we obtain a SFR $>100$ M$_\odot$ yr$^{-1}$
for SXDF-NB1006-2 (Figure~S7). On the other hand, the $J$ band
(i.e. rest-frame $\approx1500$ \AA) luminosity of this galaxy
indicates a SFR $\sim10$ M$_\odot$ yr$^{-1}$ with a standard FUV--SFR
conversion \cite{kennicutt98}. This conversion assumes a constant SFR
more than a few 100 Myr, while it actually depends on the duration 
of star formation. If the star formation age is $\sim1$ Myr, we indeed
obtain $\sim100$ M$_\odot$ yr$^{-1}$ which is consistent with the
estimation based on the [O~{\sc iii}] line. This suggests that the
target galaxy is in a young violent star formation phase.

\begin{figure}
 \begin{center}
  \includegraphics[width=8cm]{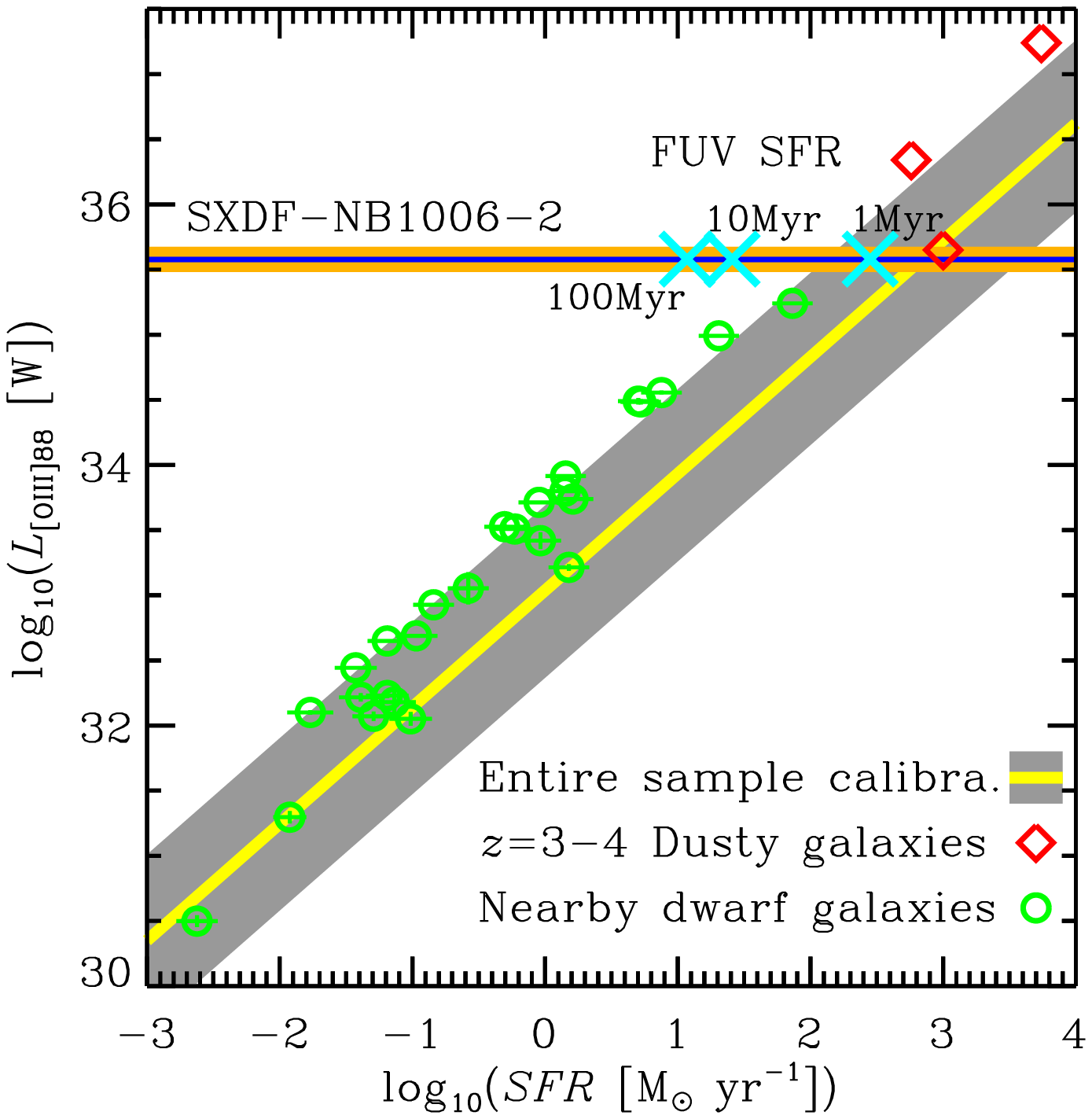}
  \vspace{0.5cm}\\
  \begin{minipage}{14cm}
   Figure~S7: \textsf{Empirical relation between the SFR and
   the [O~{\sc iii}] 88 $\mu$m line luminosity \cite{delooze14}.} 
   The best-fit relation with $\pm1\sigma$ standard deviation for a
   large compilation of the data of various kinds of galaxies including
   low-metallicity nearby dwarf galaxies (circles) and $z\sim3$--4 dusty
   starburst galaxies (diamonds) is shown by the yellow line with the
   gray shade. The [O~{\sc iii}] line luminosity with $\pm1\sigma$ 
   uncertainty of SXDF-NB1006-2 at $z=7.2$ is shown by the horizontal
   blue line with the orange shade. The crosses show SFRs estimated from
   the observed FUV luminosity of the target galaxy, under various
   assumptions on the duration of star formation. All the SFRs are
   calibrated to ones with the Salpeter IMF \cite{salpeter55} with
   0.1--100 M$_\odot$.
  \end{minipage}
 \end{center}
\end{figure}

\section{Spectral energy distribution modeling}

In order to derive physical properties of the galaxy, SXDF-NB1006-2, 
we have performed a spectral energy distribution (SED) fitting
\cite{sawicki98,ono10}. 
This is based on a standard $\chi^2$ minimization method: 
\begin{equation}
 \chi^2 = \sum_{i=1}^N \left(
  \frac{F_{i,{\rm model}}-F_{i,{\rm obs}}}{\sigma_{i,{\rm obs}}}
  \right)^2 \,,
\end{equation}
where $F_{i,{\rm model}}$, $F_{i,{\rm obs}}$, and $\sigma_{i,{\rm obs}}$
are the model flux density (or flux), the observed flux density (or
flux) and the observed uncertainty of $i$th data point, respectively. We
have used not only the broadband photometric data ($J$, $H$, $K$, $IRAC3.6$
and $IRAC4.5$) but also the narrowband $NB1006$ photometry, the [O~{\sc
iii}] 88 $\mu$m line flux and the total IR flux upper limit as
constraints. For non-detection bands and the IR flux, we simply set 
$F_{i,{\rm obs}}=0$ and take their $3\sigma$ limits as 
$\sigma_{i,{\rm obs}}$. This treatment makes the fitting favor
$F_{i,{\rm model}}$ below the $3\sigma$ limits for the non-detection
data. There are other choices to manage the non-detection data, for
example, taking $F_{i,{\rm obs}}=\sigma_{i,{\rm obs}}=1.5\sigma$
limit (option 2 of \cite{bolzonella00}) or a modification
of equation~(2) to treat the upper limits \cite{sawicki12}. We have
tried these two methods and found that the best-fit parameters do not
change but their $1\sigma$ ranges tend to be smaller. This is because
the latter two methods put a larger weight on the non-detection data.
Thus, our approach above is more conservative than the latter two
methods. The $z'$ band which is severely affected by the intergalactic 
attenuation is omitted because we have fixed the redshift to that of the
[O~{\sc iii}] line ($z=7.212$) and the non-detection in the $z'$ band
does not have much information. Therefore, the number of the constraints
is $N=8$.

\subsection{Stellar continuum}

We have adopted theoretical spectra generated with a public stellar
population synthesis code {\sc PEGASE ver.~2} \cite{pegase}.
We assume metallicities of $Z=0.0004$, 0.001, 0.002, 0.004, 0.008, 0.02,
and 0.05 with a classical solar metallicity of $Z=0.02$
\cite{anders89}. The stellar IMF is assumed to be a standard Salpeter
one \cite{salpeter55} with the range of 0.1--100 M$_\odot$. A constant
star formation history is also assumed for simplicity. In this
case, the obtained age and stellar mass are regarded as those of the
most recent star formation episode. If the galaxy has previous star
formation episodes, the true age and stellar mass are larger than those
obtained here. On the other hand, for instantaneous quantities such
as the SFR and dust attenuation, the assumption of a constant SFR is
valid in the sense of an average during the star formation episode. 
We have set a lower limit of 1 Myr in the age. 
Metallicity evolution, gas infall, outflow, nebular emission,
and dust extinction have not been considered at this stage.

\subsection{Nebular emission}

The spectra of young star-forming galaxies are significantly affected by
emission from ionized gas, so-called nebular emission \cite{schaerer09}.
We have added the nebular continuum (two-photon, bound-free, and
free-free continua) and 119 UV-to-optical ($\lambda<1$ $\mu$m in
the source rest-frame) emission lines to the model spectra following
the prescription of \cite{inoue10,inoue11}. This emission line
model is based on a large set of calculations of H~{\sc ii} regions
using a public photoionization code, {\sc cloudy} \cite{ferland13} and
reproduces the observed strengths of several prominent emission lines
such as [O~{\sc iii}] $\lambda$5007 relative to the hydrogen H$\beta$
line very well. For the [O~{\sc iii}] 88 $\mu$m line, the model
presented in \cite{inoue14} is adopted. This [O~{\sc iii}] line
model is also made with {\sc cloudy} and agrees with the available
observations of the [O~{\sc iii}] 88 $\mu$m line very well (Figure~S8).

\begin{figure}
 \begin{center}
  \includegraphics[width=8cm]{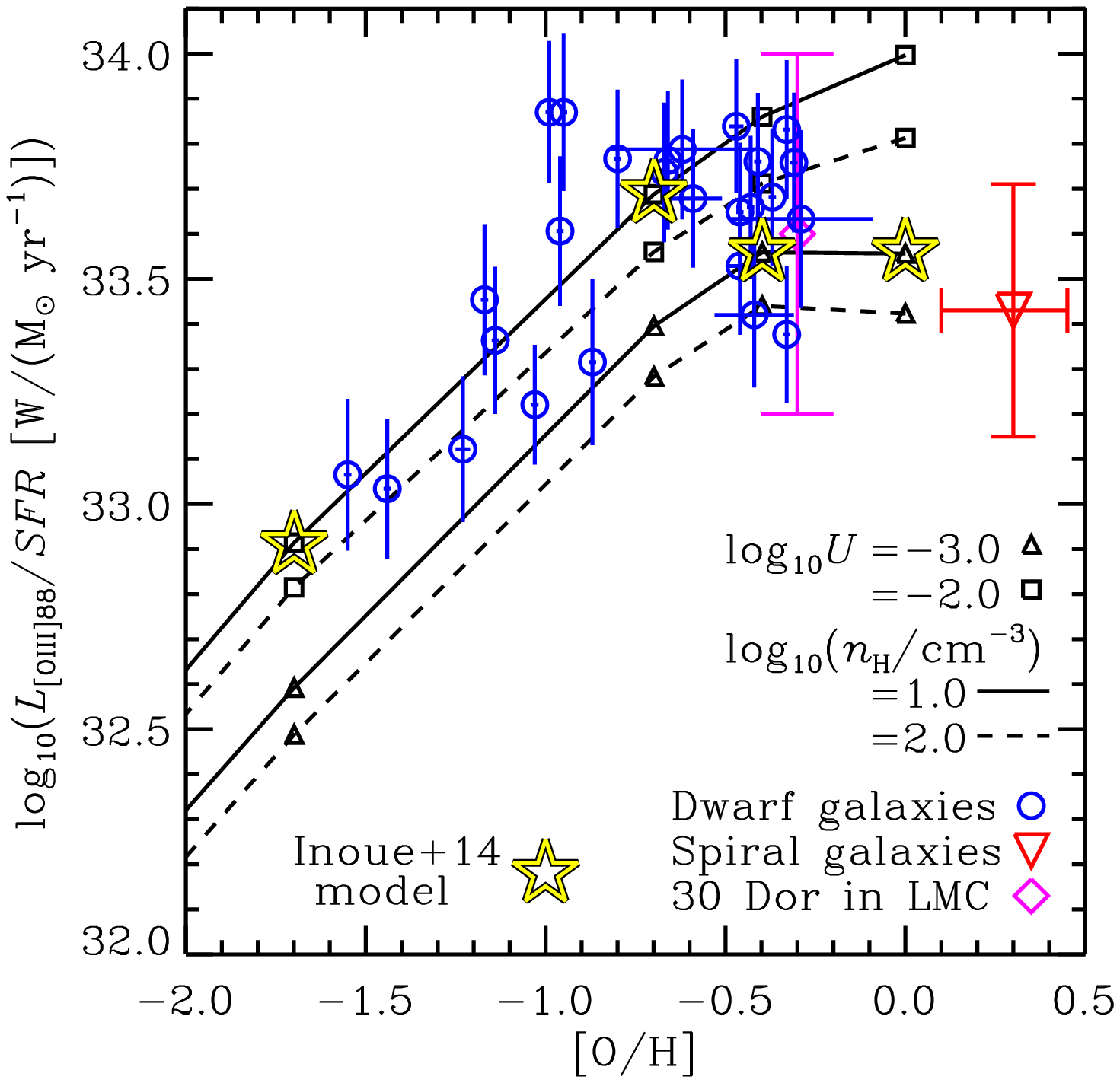}
  \vspace{0.5cm}\\
  \begin{minipage}{14cm}
   Figure~S8: \textsf{Emissivity of the [O~{\sc iii}] 88 $\mu$m line per
   a unit SFR as a function of the oxygen abundance.} The horizontal
   axis is the oxygen abundance relative to that of the Sun 
   ($[{\rm O/H}]=\log_{10}(n_{\rm O}/n_{\rm H})-\log_{10}(n_{\rm
   O}/n_{\rm H})_\odot$ with $12+\log_{10}(n_{\rm O}/n_{\rm
   H})_\odot=8.69$) \cite{asplund09}. The data of nearby dwarf galaxies
   come from observations with the {\it Herschel} satellite 
   \cite{madden13,delooze14,cormier15}. The data of LMC and nearby spiral
   galaxies as well as model predictions are taken from \cite{inoue14}. 
   We show the models with the ionization parameter $\log_{10}U$ and the
   hydrogen atom density $\log_{10}n_{\rm H}$ noted in the panel. The
   SFR is calibrated to ones with the Salpeter IMF with
   0.1--100 M$_\odot$.
  \end{minipage}
 \end{center}
\end{figure}

We allow the escape of hydrogen ionizing photons (wavelength
$\lambda<912$ \AA\ in the source rest-frame; Lyman continuum) from 
H~{\sc ii} regions and the surrounding ISM in the galaxy to the IGM. Not
only the stellar ionizing photons but also nebular bound-free ionizing
photons can escape to the IGM \cite{inoue10}. We assume that both escapes
happen with the same escape fraction, $f_{\rm esc}$, defined as the
number fraction of the escaped photons among the produced photons.

\subsection{Dust attenuation and IR luminosity}

We assume the Calzetti attenuation law \cite{calzetti94,calzetti01}
for the attenuation by dust particles in the ISM as a standard manner
found in literature, while recent studies may suggest deviations 
from the Calzetti law in high-$z$ galaxies \cite{capak15,reddy15}. 
Fortunately, the obtained attenuation amount for the target galaxy is
small and the shape of the attenuation law would not affect the
conclusions of this paper. 

The Calzetti law predicts about a factor of 2 higher attenuation for the
nebular emission than for the stellar continuum. This is because 
H~{\sc ii} regions producing the nebular emission are more deeply
embeded in gas and dust clouds than stars observed in the UV
wavelength. On the other hand, such a difference between nebular and
stellar emissions may not be supported by observations of young 
star-forming galaxies \cite{kashino13,hashimoto15}. We then introduce a
parameter, $R_{\rm gs}\equiv (E_{B-V})_{\rm gas}/(E_{B-V})_{\rm star}$,
to describe this effect and assume $R_{\rm gs}=1$ or 2. The original
Calzetti law predicts $R_{\rm gs}=2.3$.

We assume that the radiation energy attenuated by dust is finally
absorbed by dust in the ISM and thermally re-emitted in
the IR (i.e., we assume the energy scattered out to the IGM to be
negligible). This energy is compared to the total IR luminosity
estimated with the dust temperature of 40 K (Table~1), assuming the
total dust emission comes only from the star-forming regions of
interest.

\subsection{Ly$\alpha$ emission line}

The narrowband $NB1006$ photometry is mainly determined by the
Ly$\alpha$ emission line although it also contains information of the UV
continuum. Since Ly$\alpha$ photons suffer from resonant scattering 
by neutral hydrogen, the transfer in the ISM is complex. In addition to
the ISM, Ly$\alpha$ photons are also scattered by neutral hydrogen in
the IGM. Therefore, the observed Ly$\alpha$ flux becomes $F_{\rm
Ly\alpha}^{\rm obs}=F_{\rm Ly\alpha}^{\rm int}T_{\rm Ly\alpha}^{\rm
IGM}e^{-\tau^{\rm ISM}_{\rm Ly\alpha}}$, where 
$F_{\rm Ly\alpha}^{\rm int}$ is the intrinsic Ly$\alpha$ flux, 
$\tau_{\rm Ly\alpha}^{\rm ISM}$ and $T_{\rm Ly\alpha}^{\rm IGM}$ are,
respectively, the ISM optical depth and the IGM transmission for
Ly$\alpha$ photons. Recent studies of LAEs suggest that a simple recipe
like the Calzetti law with $R_{\rm gs}\simeq1$ 
(i.e. $(E_{B-V})_{\rm gas}=(E_{B-V})_{\rm stars}$) reasonably
explains the Ly$\alpha$ optical depth in the ISM inferred from the
Ly$\alpha$ line profile \cite{hashimoto15}. The IGM effect is more
severe at higher redshift due to a higher neutral fraction in the
IGM. At redshift $z\simeq7.2$, the neutral fraction, $x_{\rm HI}$, in
the IGM is estimated to be $\sim0.5$ \cite{robertson15} although it
is still uncertain. According to a cosmological radiative transfer
simulation, the Ly$\alpha$ transmission through the IGM with an average
$x_{\rm HI}=0.5$ is $T_{\rm Ly\alpha}^{\rm IGM}=0.35_{-0.15}^{+0.10}$
\cite{jensen13}. Since this $T_{\rm Ly\alpha}^{\rm IGM}$ range
also covers a wide range of $x_{\rm HI}=0.2$--0.8 \cite{jensen13}, 
we assume that the $T_{\rm Ly\alpha}^{\rm IGM}$ range encloses the
uncertainty of $x_{\rm HI}$.

\subsection{Results}

We use $N=8$ observational constraints: 6 photometric data ($NB1006$,
$J$, $H$, $K$, $IRAC3.6$ and $IRAC4.5$) and the [O~{\sc iii}] 88 $\mu$m
line flux and the total IR flux. On the other hand, there are 7 model
parameters: the metallicity $Z$, the IGM Ly$\alpha$ transmission 
$T_{\rm Ly\alpha}^{\rm IGM}$, the dust attenuation ratio of nebular to
stellar emissions $R_{\rm gs}$, the SFR, the age, the stellar dust
attenuation $(E_{B-V})_{\rm star}$, and the escape fraction of ionizing
photons $f_{\rm esc}$. For the first 3 parameters, we fixed the values:
$Z=0.0004$, 0.001, 0.002, 0.004, 0.008, 0.02($=Z_\odot)$, or 0.05, 
$T_{\rm Ly\alpha}^{\rm IGM}=0.20$, 0.30, 0.35, 0.40, or 0.45,
and $R_{\rm gs}=1$ or 2. We then searched for the best set of the rest 4
parameters by a standard $\chi^2$ method. The resultant best-fit values
and their 68.4\% ranges (i.e. $\Delta \chi^2<1$) are summarized in
Table~S4, where we only show the cases with $R_{\rm gs}=1$ but the
$R_{\rm gs}=2$ cases are not very different because of a very small dust
attenuation.

We find that the minimum $\chi^2$ value is obtained with the metallicity
$Z=0.002(=0.1Z_\odot)$ but the $0.001\leq Z\leq0.02$ cases give equally
good fit results. On the other hand, the $Z=0.0004$  and 0.05 cases
are rejected at a $>95\%$ confidence level, except for the case of
$Z=0.0004$ and $T_{\rm Ly\alpha}^{\rm IGM}=0.20$ which can be rejected
at a $\sim90\%$. Therefore, the galaxy, SXDF-NB1006-2, is likely to
have a metallicity of $0.05\leq Z/Z_\odot\leq1$.

The best-fit model is $Z=0.002$, $T_{\rm Ly\alpha}^{\rm IGM}=0.40$, 
$\log_{10}(SFR/{\rm M}_\odot~{\rm yr}^{-1})=2.54$, the age of 1 Myr, 
no dust attenuation, and $f_{\rm esc}=0.54$, which is shown in
Figures~2B--2D. The obtained 1 Myr age is in fact the lower limit of the
population synthesis model. This shortest age is favored by the very
blue UV color of the galaxy ($J-H<-0.18$ corresponding to the UV slope
$\beta<-2.6$ [$3\sigma$]). This blue UV color also favors small dust
attenuation but the upper limit of the dust IR luminosity gives a
stronger constraint on the dust attenuation (see Figure~S9 and a
discussion below). If we fix $Z$ and  
$T_{\rm Ly\alpha}^{\rm IGM}\geq0.30$, a non-zero $f_{\rm esc}$ tends to
be favored. However, there are many sets of $Z$ and 
$T_{\rm Ly\alpha}^{\rm IGM}$ giving $\chi^2$ as good as the best-fit
case statistically. We derive the 68.4\% ranges of each parameters
(i.e. $\Delta \chi^2<1$) among all cases examined with $R_{\rm gs}=1$.
The results are as follows:
$1.83\leq \log_{10}(SFR~[{\rm M_\odot~yr^{-1}}])\leq2.71$, 
$6.00\leq \log_{10}(t~[{\rm yr}]) \leq7.00$, 
$0.00\leq (E_{B-V})_{\rm star}<0.04$, and 
$0.00\leq f_{\rm esc} \leq 0.71$. For the stellar mass, we find
$\log_{10}(M_{\rm star}/{\rm M}_\odot)=8.53$ as the best-fit and the 
68.4\% range of  
$8.32\leq \log_{10}(M_{\rm star}/{\rm M}_\odot)<9.33$
as a joint constraint of $\log_{10}(SFR~[{\rm M_\odot~yr^{-1}}])$ 
and $\log_{10}(t~[{\rm yr}])$ (i.e. $\Delta \chi^2<2.3$).

\begin{table}
 \begin{center}
  Table~S4: \textsf{A summary of SED fitting results.}
  \small
 \begin{tabular}{ccccccc}
  \hline \hline
  $Z$ & $T_{\rm Ly\alpha}^{\rm IGM}$ & 
  log$_{10}(SFR~{\rm [M_\odot~yr^{-1}]})$ & log$_{10}(t~{\rm [yr]})$ 
  & $(E_{B-V})_{\rm star}$ & $f_{\rm esc}$ & $\chi^2_{\rm min}$ \\
  \hline
0.0004 & 0.45 & $2.83_{-0.12}^{+0.10}$ & $6.00_{-0.00}^{+0.00}$
 & $0.08_{-0.03}^{+0.02}$ & $0.44_{-0.16}^{+0.14}$ & 9.376 \\
0.0004 & 0.40 & $2.83_{-0.14}^{+0.09}$ & $6.00_{-0.00}^{+0.00}$
 & $0.08_{-0.03}^{+0.01}$ & $0.40_{-0.18}^{+0.14}$ & 8.417 \\
0.0004 & 0.35 & $2.79_{-0.12}^{+0.11}$ & $6.00_{-0.00}^{+0.00}$
 & $0.07_{-0.02}^{+0.02}$ & $0.34_{-0.17}^{+0.15}$ & 7.331 \\  
0.0004 & 0.30 & $2.75_{-0.12}^{+0.13}$ & $6.00_{-0.00}^{+0.00}$
 & $0.06_{-0.02}^{+0.02}$ & $0.28_{-0.18}^{+0.16}$ & 6.258 \\  
0.0004 & 0.20 & $2.65_{-0.11}^{+0.15}$ & $6.00_{-0.00}^{+0.00}$
 & $0.04_{-0.02}^{+0.02}$ & $0.09_{-0.09}^{+0.20}$ & 4.035 \\  
  \hline
0.0010 & 0.45 & $2.74_{-0.15}^{+0.10}$ & $6.00_{-0.00}^{+0.00}$
 & $0.05_{-0.03}^{+0.01}$ & $0.52_{-0.14}^{+0.10}$ & 3.304 \\
0.0010 & 0.40 & $2.70_{-0.14}^{+0.11}$ & $6.00_{-0.00}^{+0.00}$
 & $0.04_{-0.02}^{+0.02}$ & $0.49_{-0.15}^{+0.10}$ & 2.855 \\
0.0010 & 0.35 & $2.66_{-0.14}^{+0.13}$ & $6.00_{-0.00}^{+0.00}$
 & $0.03_{-0.02}^{+0.02}$ & $0.44_{-0.15}^{+0.13}$ & 2.481 \\
0.0010 & 0.30 & $2.61_{-0.14}^{+0.15}$ & $6.00_{-0.00}^{+0.00}$
 & $0.02_{-0.02}^{+0.02}$ & $0.38_{-0.15}^{+0.15}$ & 2.166 \\
0.0010 & 0.20 & $2.53_{-0.25}^{+0.14}$ & $6.00_{-0.00}^{+0.30}$
 & $0.00_{-0.00}^{+0.03}$ & $0.26_{-0.26}^{+0.16}$ & 1.871 \\
  \hline
0.0020 & 0.45 & $2.58_{-0.18}^{+0.14}$ & $6.00_{-0.00}^{+0.30}$
 & $0.01_{-0.01}^{+0.03}$ & $0.57_{-0.14}^{+0.11}$ & 1.695 \\
0.0020 & 0.40 & $2.54_{-0.20}^{+0.14}$ & $6.00_{-0.00}^{+0.30}$
 & $0.00_{-0.00}^{+0.04}$ & $0.54_{-0.18}^{+0.12}$ & {\bf 1.629} \\
0.0020 & 0.35 & $2.54_{-0.28}^{+0.13}$ & $6.00_{-0.00}^{+0.30}$
 & $0.00_{-0.00}^{+0.03}$ & $0.52_{-0.25}^{+0.11}$ & 1.694 \\
0.0020 & 0.30 & $2.34_{-0.25}^{+0.30}$ & $6.30_{-0.30}^{+0.30}$
 & $0.01_{-0.01}^{+0.03}$ & $0.34_{-0.29}^{+0.27}$ & 1.890 \\
0.0020 & 0.20 & $2.09_{-0.11}^{+0.31}$ & $6.48_{-0.18}^{+0.22}$
 & $0.00_{-0.00}^{+0.02}$ & $0.02_{-0.02}^{+0.38}$ & 2.075 \\
  \hline
0.0040 & 0.45 & $2.30_{-0.30}^{+0.29}$ & $6.30_{-0.30}^{+0.40}$
 & $0.00_{-0.00}^{+0.03}$ & $0.50_{-0.30}^{+0.21}$ & 1.701 \\
0.0040 & 0.40 & $2.31_{-0.46}^{+0.12}$ & $6.30_{-0.00}^{+0.70}$
 & $0.00_{-0.00}^{+0.03}$ & $0.50_{-0.50}^{+0.11}$ & 1.710 \\
0.0040 & 0.35 & $2.11_{-0.30}^{+0.29}$ & $6.48_{-0.18}^{+0.60}$
 & $0.00_{-0.00}^{+0.03}$ & $0.31_{-0.31}^{+0.28}$ & 1.875 \\
0.0040 & 0.30 & $2.11_{-0.30}^{+0.25}$ & $6.48_{-0.18}^{+0.52}$
 & $0.00_{-0.00}^{+0.02}$ & $0.28_{-0.28}^{+0.27}$ & 1.955 \\
0.0040 & 0.20 & $1.94_{-0.10}^{+0.24}$ & $6.70_{-0.22}^{+0.26}$
 & $0.00_{-0.00}^{+0.01}$ & $0.03_{-0.03}^{+0.34}$ & 2.981 \\
  \hline
0.0080 & 0.45 & $2.31_{-0.30}^{+0.30}$ & $6.30_{-0.30}^{+0.40}$
 & $0.01_{-0.01}^{+0.03}$ & $0.43_{-0.35}^{+0.25}$ & 1.847 \\
0.0080 & 0.40 & $2.27_{-0.34}^{+0.31}$ & $6.30_{-0.30}^{+0.48}$
 & $0.00_{-0.00}^{+0.03}$ & $0.39_{-0.39}^{+0.26}$ & 1.779 \\
0.0080 & 0.35 & $2.27_{-0.37}^{+0.27}$ & $6.30_{-0.30}^{+0.54}$
 & $0.00_{-0.00}^{+0.03}$ & $0.37_{-0.37}^{+0.23}$ & 1.790 \\
0.0080 & 0.30 & $2.07_{-0.17}^{+0.29}$ & $6.48_{-0.18}^{+0.30}$
 & $0.00_{-0.00}^{+0.02}$ & $0.13_{-0.13}^{+0.34}$ & 1.954 \\
0.0080 & 0.20 & $1.99_{-0.07}^{+0.32}$ & $6.60_{-0.30}^{+0.10}$
 & $0.00_{-0.00}^{+0.01}$ & $0.00_{-0.00}^{+0.40}$ & 2.974 \\
  \hline
0.0200 & 0.45 & $2.44_{-0.38}^{+0.12}$ & $6.00_{-0.00}^{+0.48}$
 & $0.00_{-0.00}^{+0.03}$ & $0.52_{-0.43}^{+0.11}$ & 1.667 \\
0.0200 & 0.40 & $2.45_{-0.50}^{+0.09}$ & $6.00_{-0.00}^{+0.70}$
 & $0.00_{-0.00}^{+0.03}$ & $0.51_{-0.51}^{+0.10}$ & 1.845 \\
0.0200 & 0.35 & $2.20_{-0.27}^{+0.30}$ & $6.30_{-0.30}^{+0.40}$
 & $0.00_{-0.00}^{+0.03}$ & $0.26_{-0.26}^{+0.32}$ & 1.771 \\
0.0200 & 0.30 & $2.02_{-0.08}^{+0.45}$ & $6.48_{-0.48}^{+0.22}$
 & $0.00_{-0.00}^{+0.02}$ & $0.01_{-0.01}^{+0.52}$ & 1.937 \\
0.0200 & 0.20 & $2.04_{-0.07}^{+0.22}$ & $6.48_{-0.18}^{+0.12}$
 & $0.00_{-0.00}^{+0.01}$ & $0.00_{-0.00}^{+0.32}$ & 3.001 \\
  \hline
0.0500 & 0.45 & $2.60_{-0.11}^{+0.10}$ & $6.00_{-0.00}^{+0.00}$
 & $0.08_{-0.03}^{+0.02}$ & $0.79_{-0.06}^{+0.05}$ & 13.145 \\
0.0500 & 0.40 & $2.60_{-0.12}^{+0.10}$ & $6.00_{-0.00}^{+0.00}$
 & $0.08_{-0.03}^{+0.02}$ & $0.77_{-0.08}^{+0.06}$ & 12.067 \\
0.0500 & 0.35 & $2.60_{-0.12}^{+0.09}$ & $6.00_{-0.00}^{+0.00}$
 & $0.08_{-0.03}^{+0.02}$ & $0.74_{-0.08}^{+0.07}$ & 10.855 \\
0.0500 & 0.30 & $2.57_{-0.10}^{+0.11}$ & $6.00_{-0.00}^{+0.00}$
 & $0.07_{-0.02}^{+0.02}$ & $0.71_{-0.09}^{+0.07}$ &  9.498 \\
0.0500 & 0.20 & $2.54_{-0.12}^{+0.09}$ & $6.00_{-0.00}^{+0.00}$
 & $0.06_{-0.02}^{+0.02}$ & $0.62_{-0.10}^{+0.08}$ &  6.311 \\
  \hline
  \end{tabular}
 \end{center}
\end{table}

\begin{figure}
 \begin{center}
  \includegraphics[width=7cm]{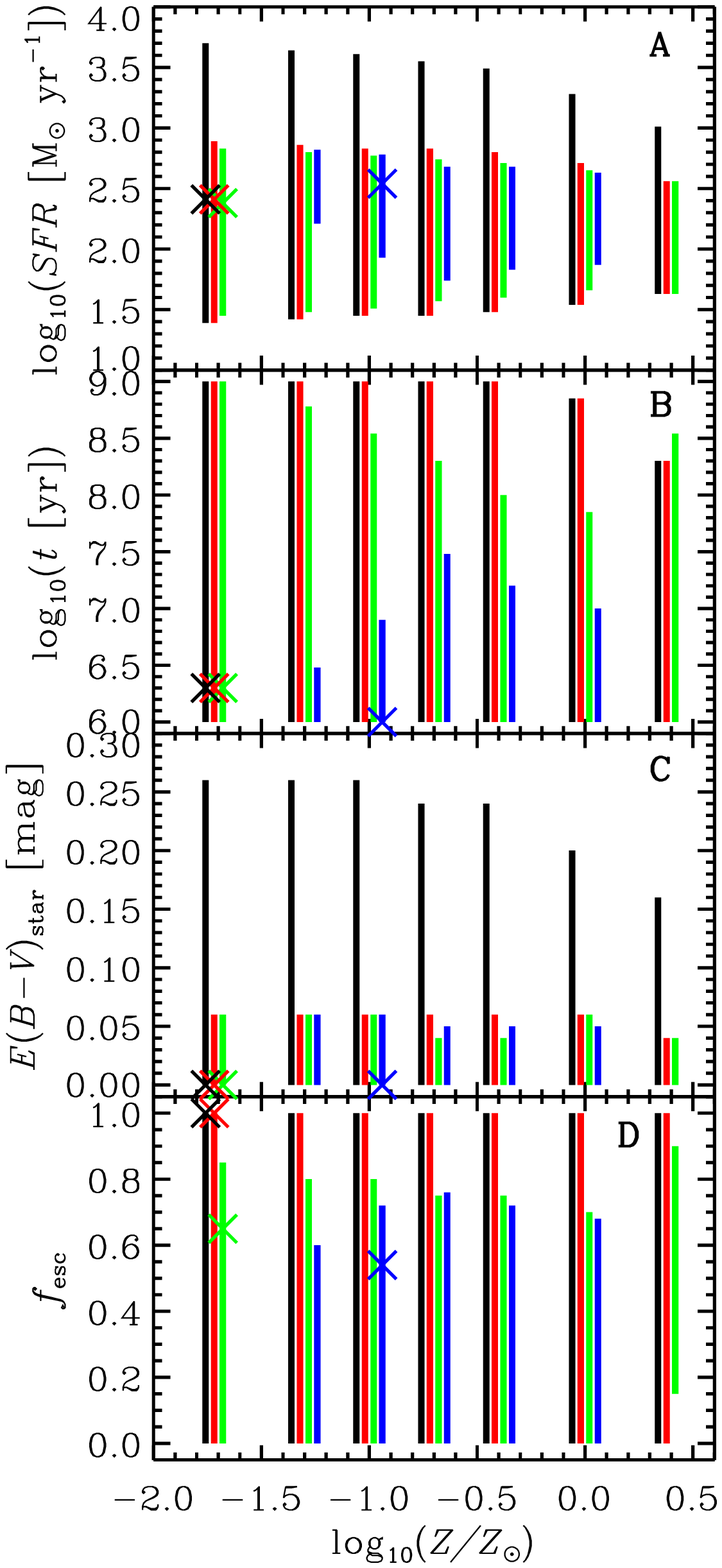}
  \vspace{0.5cm}\\
  \begin{minipage}{14cm}
   Figure~S9: \textsf{Visualizations of SED fitting results.} 
   (A) The SFR ranges within $\Delta \chi^2<2.30$ (a $\pm1\sigma$ range
   for two parameters) from the best-fit models whose locations are
   shown by the crosses. The black, red, green and blue colors represent
   the cases using only broadband (BB) photometries ($J$, $H$, $K$,
   $IRAC3.6$, and $IRAC4.5$), BB + the infrared (IR) luminosity limit,
   BB + IR + the narrowband (NB) photometry ($NB1006$), and BB + IR + NB
   + the [O~{\sc iii}] line, respectively. The most metal-poor and
   metal-rich models were rejected in the last case. (B) The same as A
   but for the star formation age. (C) The same as A but for the dust
   attenuation for stars. (D) The same as A but for the escape
   fraction of ionizing photons.
  \end{minipage}
 \end{center}
\end{figure}

The [O~{\sc iii}] 88 $\mu$m line flux is the most powerful constraint in
the SED modeling of SXDF-NB1006-2 (Figure~S9). Without the [O~{\sc
iii}] line, we cannot obtain any meaningful constraint on the
metallicity, whereas using the [O~{\sc iii}] line, we can reject the
most metal-poor and metal-rich cases examined (i.e. $Z=0.0004$
and 0.05). The [O~{\sc iii}] line also improves the constraints on
other parameters dramatically, although the best-fit values of the
SFR, age, and dust attenuation are not very different regardless of the
sets of the data used in the fitting. Generally, less data give a weaker
convergence around the best-fit values as expected. When we use only the
broadband data as constraints, only a few models are rejected. The
IR luminosity limit greatly improves the constraint on the dust
attenuation (Figure~S9C). Using the narrowband $NB1006$ which
includes the Ly$\alpha$ line slightly improves overall constraints.

\section{Comparison with a cosmological simulation}

We now compare SXDF-NB1006-2 with galaxies in a large
cosmological hydrodynamic simulation of galaxy formation and evolution
\cite{shimizu14,shimizu15} which reproduces the observed UV
luminosity functions and UV colors of Lyman break galaxies at
$z\sim7$--10 very well. Comparing SXDF-NB1006-2 with the galaxies taken
from the simulation output at z=7.22, we discuss implications on the
physical and chemical properties of the target galaxy. A brief
explanation of the simulation is as follows, while further details of the
simulation are found in \cite{shimizu14,shimizu15}. The simulation code
we used is {\sc gadget-3}: an updated version of {\sc gadget-2}
\cite{springel05}. The physical recipes describing the star formation,
chemical evolution, supernovae and radiation feedback
\cite{okamoto08,okamoto10,okamoto14} are implemented. We employ
$2\times640^3$ particles for dark matter and gas in a comoving volume of
$100h^{-1}$ Mpc cube. The mass of a dark matter particle is
$2.84\times10^8h^{-1}$ M$_\odot$ and the mass of a gas particle is
initially $5.17\times10^7h^{-1}$ M$_\odot$. The softening length for the
gravitational force is set to be $6.0 h^{-1}$ comoving kpc. The gas
particles may form star particles if the star formation criteria are
satisfied. We also implement the emission line model
\cite{inoue11,inoue14} into the simulation, assuming the zero escape of
ionizing photons.

\begin{figure}
 \begin{center}
  \includegraphics[width=12cm]{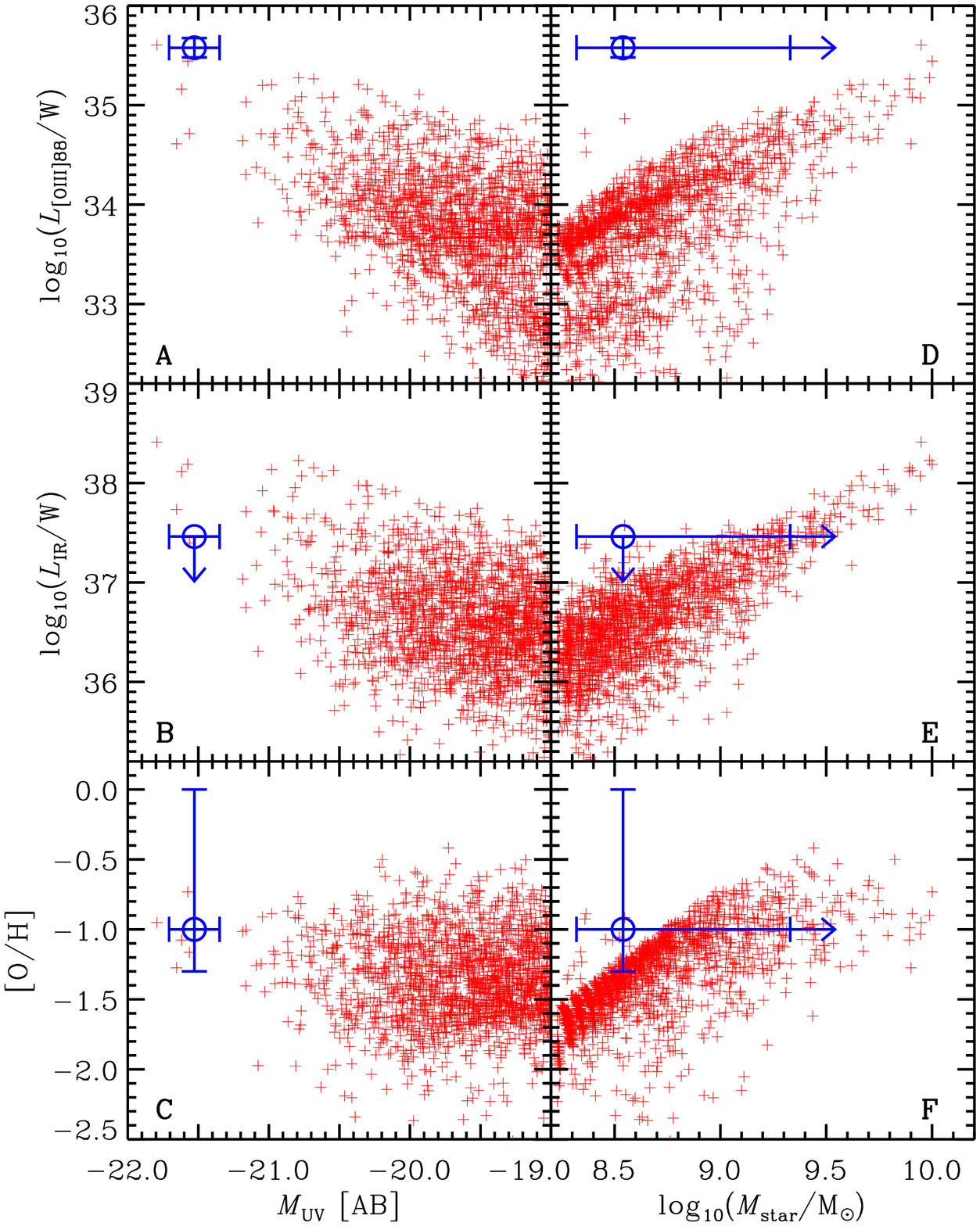}
  \vspace{0.5cm}\\
  \begin{minipage}{14cm}
   Figure~S10: \textsf{Comparisons of SXDF-NB1006-2 with galaxies in a
   cosmological hydrodynamic simulation.} Panels A--C show the [O~{\sc
   iii}] 88 $\mu$m line luminosity, the dust IR luminosity, and the
   oxygen abundance (or metallicity), respectively, as a function of the
   absolute UV magnitude uncorrected for the dust attenuation. Panels
   D--F are the same as A--C but as a function of the stellar mass. The
   circles with error-bars are the data of SXDF-NB1006-2; we take a
   $3\sigma$ upper limit for the dust IR luminosity with an dust
   temperature of 40 K and an emissivity index of 1.5. The stellar
   mass obtained from the SED fitting should be regarded as a lower
   limit because of our simple constant star formation history. The plus
   marks are galaxies at $z=7.22$ taken from the simulation
   \cite{shimizu15}.
  \end{minipage}
 \end{center}
\end{figure}

Comparisons of SXDF-NB1006-2 with the $z=7.22$ galaxies taken from
the simulation (Figure~S10) show that there are five galaxies with
similar UV luminosities to that of SXDF-NB1006-2 (panels A--C). The
[O~{\sc iii}] line luminosity of SXDF-NB1006-2 is very close to the two
highest ones among the five. The SFRs of the two simulated galaxies are
51 and 92 M$_\odot$ yr$^{-1}$, whereas the SFR of SXDF-NB1006-2 is
estimated at $\sim300$ M$_\odot$ yr$^{-1}$ from the SED modeling. The
best-fit SED model suggests a $\sim50\%$ escape of ionizing photons,
indicating a factor of $\sim2$ reduction of the [O~{\sc iii}] line
luminosity per SFR in SXDF-NB1006-2. This partly accounts for the SFR
difference in spite of similar [O~{\sc iii}] line luminosities. In any
case, SXDF-NB1006-2 seems to be in an intense starburst phase which
enhances the [O~{\sc iii}] line luminosity.

We also find that the average of the IR luminosities of the five
simulated galaxies are 0.5 dex higher than the $3\sigma$ upper limit of
SXDF-NB1006-2 (panel B), suggesting that the target galaxy has much less
dust than the simulated galaxies. In fact, these simulated galaxies have
$(E_{B-V})_{\rm star}=0.15$, whereas that of SXDF-NB1006-2 is less than
0.04 mag from the SED fitting. On the other hand, the metallicity of
SXDF-NB1006-2 is similar to or higher than those of the five (panel C).
These indicates that SXDF-NB1006-2 has a much less dust-to-metal mass
ratio than the simulated galaxies where we have assumed the ratio to be
0.5 as in the ISM of the Milky Way \cite{draine11} and of the Solar
neighborhood \cite{kimura03}.

\begin{figure}
 \begin{center}
  \includegraphics[width=8cm]{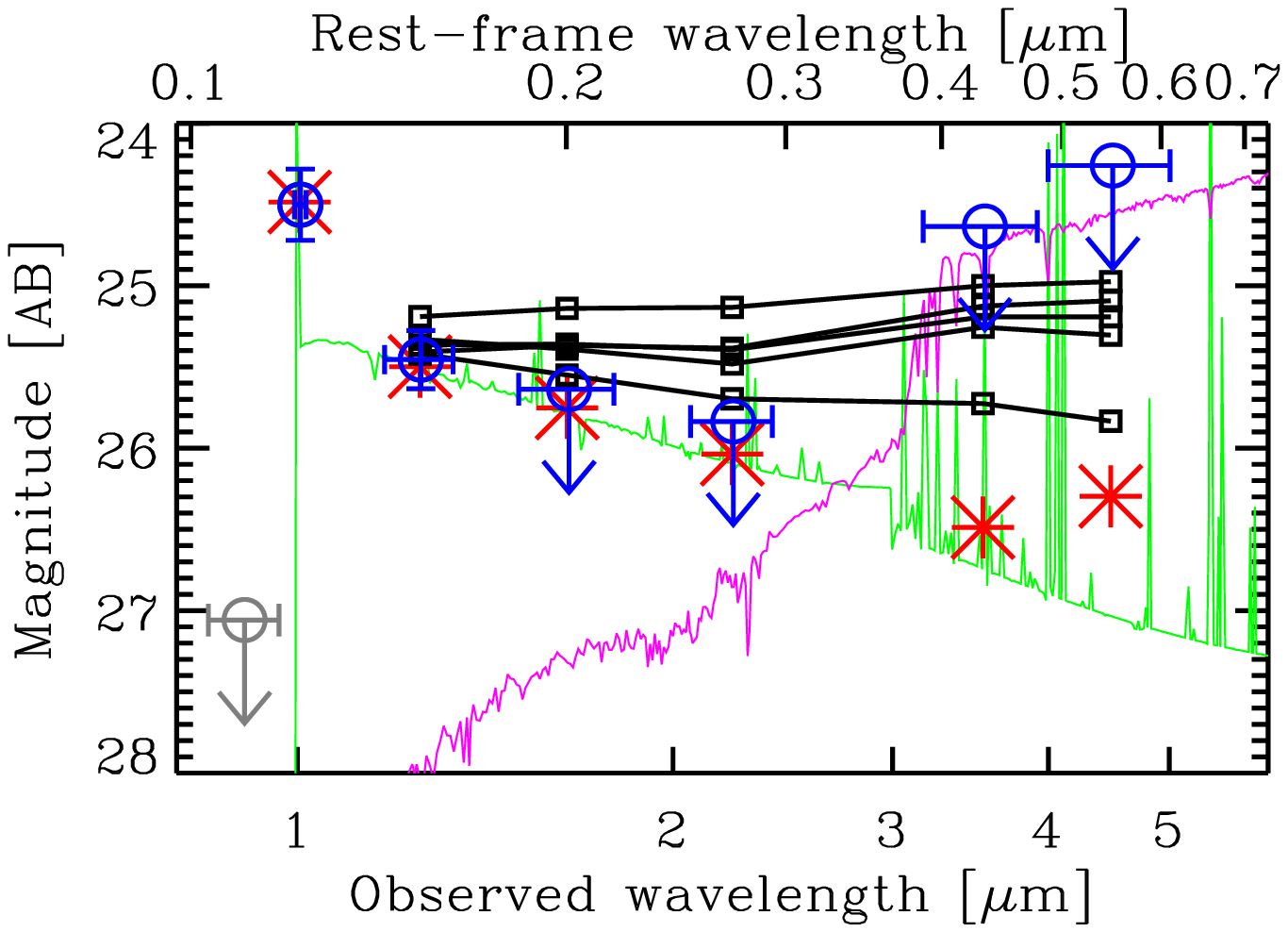}
  \vspace{0.5cm}\\
  \begin{minipage}{12cm}
   Figure~S11: \textsf{Comparison of spectral energy distributions.}
   The circles with the error-bars are the data of SXDF-NB1006-2. The
   green solid line is the best-fit model and the asterisks are those
   convolved with the filter response curves. The magenta solid line
   shows a maximally possible passive stellar population
   ($M_*=5\times10^{10}$ M$_\odot$ and age of 700 Myr). The squares
   connected with lines are the five simulated galaxies having a 
   similar UV luminosity to SXDF-NB1006-2.
  \end{minipage}
 \end{center}
\end{figure}

The stellar mass of SXDF-NB1006-2 obtained from the SED fitting is an
order of magnitude smaller than those expected in the simulation (panels
D--F). This small stellar mass is mainly due to the very short age
($\sim1$ Myr) which is constrained from the blue UV color of the galaxy.
Indeed, the observed UV color of SXDF-NB1006-2 is much bluer than those
of the five simulated galaxies (Figure~S11). However, it is
still possible that in addition to the $\sim1$ Myr starburst,
SXDF-NB1006-2 has an underlying passive stellar population with a
stellar mass of $<5\times10^{10}$ M$_\odot$ and an age of 700 Myr
($\approx$ the age of the Universe at $z=7.2$) without violating
the Spitzer data at 3.6 $\mu$m and 4.5 $\mu$m (Figure~S11). Even in  
this case, the blue UV color of SXDF-NB1006-2 is not affected by this
passive stellar population and the conclusion that SXDF-NB1006-2 is a
very young starburst and has little dust content is preserved.

\section{Ionizing photon emission efficiency}

There is an important quantity in the context of cosmic reionization: 
the ionizing photon injection rate into the IGM per unit UV luminosity
density of galaxies \cite{bouwens15b}. This can be expressed as  
\begin{equation}
 f_{\rm esc}\xi_{\rm ion}=f_{\rm esc}
  \left(\frac{Q_{\rm H}}{SFR}\right)
  \frac{SFR}{L_{\nu_{1500}}^{\rm obs}}\,,
\end{equation}
where $(Q_{\rm H}/SFR)$ is the intrinsic production rate of hydrogen
ionizing photons ($\lambda<912$ \AA) per a unit SFR and 
$L_{\nu_{1500}}^{\rm obs}$ is the luminosity density at UV 
($\lambda\sim1500$ \AA\ in the source rest-frame). We estimate this rate
for SXDF-NB1006-2. For the stellar population corresponding to  
the best-fit model with $Z=0.002$ and the 1 Myr age, 
$(Q_{\rm H}/SFR)=2.58\times10^{52}$ s$^{-1}$. 
Different metallicities cause only 0.03 dex variation. 
However, the star formation age affects the production rate; a constant
SFR of 10 Myr ($+1\sigma$ age) gives a 0.40 dex larger production rate.
If we adopt the best-fit stellar population and uncertainties of the SFR
and $f_{\rm esc}$ in the case of $T_{\rm Ly\alpha}^{\rm IGM}=0.40$ 
(Table~S4), we obtain 
$\log_{10}(f_{\rm esc}\xi_{\rm ion}/{\rm Hz~erg^{-1}})=25.44^{+0.18}_{-0.26}$ 
by using an error propagation formula for the logarithm. If we adopt the
uncertainties of the final estimates of the SFR and $f_{\rm esc}$
(Table~1) and take into account the 0.40 dex upward uncertainty caused
by the star formation age, we obtain 
$\log_{10}(f_{\rm esc}\xi_{\rm ion}/{\rm Hz~erg^{-1}})=25.44^{+0.46}_{-0.84}$. 

We compare the obtained ionizing photon injection rate per UV luminosity
density for SXDF-NB1006-2 with that required to reproduce the comoving
volume emissivity of ionizing photons of 
$\log_{10}(\dot N_{\rm ion} {\rm [s^{-1}~Mpc^{-3}]})=50.79\pm0.06$ at
$z\sim7$ which is estimated from various observational constraints on
cosmic reionization with a parametric expression of the 
$\dot N_{\rm ion}$ evolution \cite{bouwens15b}. Since the photon
injection rate is given by  
$\log_{10}(f_{\rm esc}\xi_{\rm ion})=
\log_{10}(\dot N_{\rm ion})-\log_{10}(\rho_{\rm UV})$, where
$\rho_{\rm UV}$ is the comoving UV luminosity density, we need to
integrate a UV luminosity function. Here we consider two UV luminosity
functions at $z\sim7$ reported by \cite{bouwens15} and
\cite{finkelstein15}. We have performed a set of Monte Carlo
realizations of Schechter function fits to the luminosity functions
fluctuated based on the quoted uncertainties and obtained a set of
$\rho_{\rm UV}$ as a function of a faint-limit of $M_{\rm UV}$ by
integrating the best-fit Schechter function in each realization. We have
also taken into account the uncertainty of the ionizing photon
emissivity, $\log_{10}(\dot N_{\rm ion})$, in the procedure to obtain
the emission efficiency, $\log_{10}(f_{\rm esc}\xi_{\rm ion})$. Finally,
we have calculated the mean and standard deviation of 
$\log_{10}(f_{\rm esc}\xi_{\rm ion})$ among the Monte Carlo
realizations as a function of the faint $M_{\rm UV}$ limit. 

From this comparison (Figure~S12), we find that it is difficult to
reproduce the ionizing photon emissivity at $z\sim7$ only by galaxies
brighter than SXDF-NB1006-2 ($M_{\rm UV}=-21.53$), even if these
galaxies emit ionizing photons as strong as that galaxy, because of the
small number density of such bright galaxies. On the other hand, if
galaxies with $M_{\rm UV}<-17$, which are already detected in deep HST
surveys, have an ionizing photon emission efficiency similar to
SXDF-NB1006-2, the ionizing photon emissivity is likely to be achieved
or even exceeded by 0.4--0.6-dex. However, if objects emitting ionizing
photons as strong as SXDF-NB1006-2 are rare among galaxies with 
$M_{\rm UV}<-17$, fainter, currently undetected galaxies should
contribute to the cosmic ionizing photon emissivity.

\begin{figure}
 \begin{center}
  \includegraphics[width=8cm]{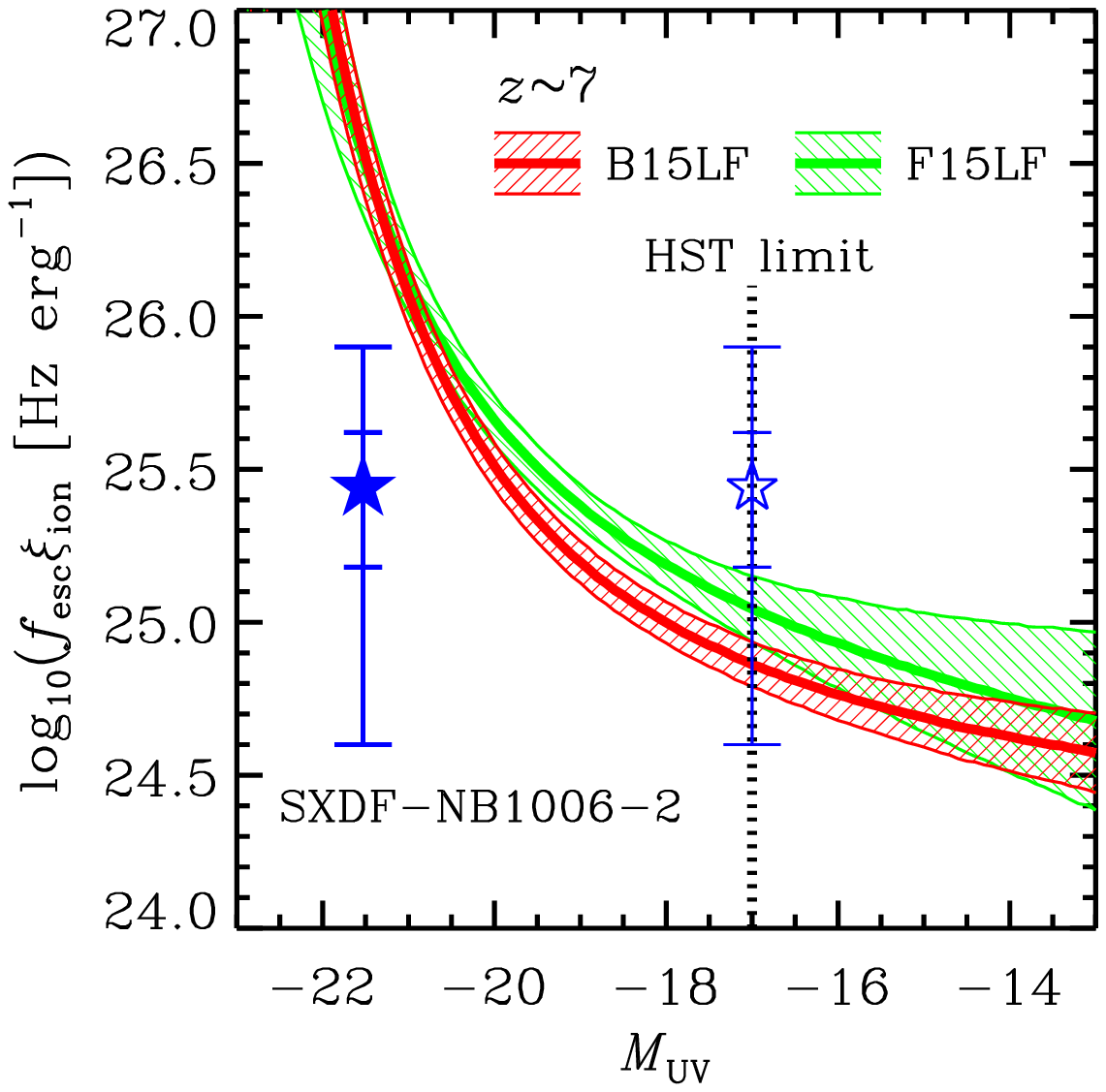}
  \vspace{0.5cm}\\
  \begin{minipage}{12cm}
   Figure~S12: \textsf{Comparison of ionizing photon injection rates
   into the IGM per UV luminosity density.} The red and green shades
   show the rates required to match the cosmic ionizing photon
   emissivity estimated at $z\sim7$ \cite{bouwens15b}, when we integrate 
   the UV luminosity function of \cite{bouwens15} (denoted as B15LF) or
   \cite{finkelstein15} (denoted as F15LF), respectively, down to the
   faint UV magnitude limit indicated on the horizontal axis. 
   The rate obtained from SXDF-NB1006-2 is shown by a five-pointed-star
   with error-bars. The smaller error-bars show the case with the
   best-fit stellar population and $T_{\rm Ly\alpha}^{\rm IGM}=0.40$,
   but the larger error-bars show the case considering all uncertainties
   in our estimates. The vertical dotted line at $M_{\rm UV}=-17$
   indicates a detection limit with HST/WFC3 \cite{bouwens15}. The open
   star with error-bars shows the data of SXDF-NB1006-2 at the HST
   detection limit for a reference.
  \end{minipage}
 \end{center}
\end{figure}

\end{document}